\documentclass[lettersize,journal]{IEEEtran}
\usepackage{cite}
\usepackage{amsmath,amssymb,amsfonts}
\usepackage{algorithmic}
\usepackage{graphicx,color}
\usepackage{textcomp}
\usepackage{xcolor}
\usepackage{hyperref}
\hypersetup{hidelinks=true}
\usepackage{algorithm,algorithmic}
\usepackage{yhmath}
\usepackage{bbding}
\usepackage{mathrsfs}
\usepackage{tabularx}

\usepackage{graphicx} 

\usepackage{epstopdf} 

\usepackage{xcolor}

\usepackage{bbm} 
\usepackage{bm}

\usepackage{algorithm}
\usepackage{url}

\usepackage{multirow} 

\usepackage{subfig}

\usepackage{tablefootnote}

\usepackage{caption}

\allowdisplaybreaks[1]

\usepackage{booktabs}
\usepackage{threeparttable}
\usepackage{footnote}
\usepackage{enumerate}

\usepackage{yhmath}

\usepackage{array}

\usepackage{diagbox}

\newcommand\bigfrown[2][\textstyle]{\ensuremath{%
  \array[b]{c}\text{\scalebox{1.7}{$#1\frown$}}\\[-1.7ex]#1#2\endarray}}

\usepackage{ntheorem}

\newtheorem{prop}{Proposition}

\begin{document}


\title{Optimizing QoE-Privacy Tradeoff for Proactive\\ VR Streaming}

\author{
	\IEEEauthorblockN{Xing Wei, Shengqian Han, Chenyang Yang, and Chengjian Sun}
	\\ \IEEEauthorblockA{Beihang University, Beijing, China\\
		Email: \{weixing, sqhan, cyyang, sunchengjian\}@buaa.edu.cn}
}

\maketitle

\begin{abstract}
Proactive virtual reality (VR) streaming requires users to upload viewpoint-related information, raising significant privacy concerns. Existing strategies preserve privacy by introducing errors to viewpoints, which, however, compromises the quality of experience (QoE) of users. In this paper, we first delve into the analysis of the viewpoint leakage probability achieved by existing privacy-preserving approaches. We determine the optimal distribution of viewpoint errors that minimizes the viewpoint leakage probability. Our analyses show that existing approaches cannot fully eliminate viewpoint leakage. Then, we propose a novel privacy-preserving approach that introduces noise to uploaded viewpoint prediction errors, which can ensure zero viewpoint leakage probability. Given the proposed approach, the tradeoff between privacy preservation and QoE is optimized to minimize the QoE loss while satisfying the privacy requirement. Simulation results validate our analysis results and demonstrate that the proposed approach offers a promising solution for balancing privacy and QoE.
\end{abstract}


\maketitle

\section{Introduction}

Virtual reality (VR) services are widely recognized as a key application in future wireless communication systems. To achieve immersive user experiences, various behavior trajectories of VR users, including viewpoint, gaze, and hand trajectories, are tracked and uploaded~\cite{metaverse_WCM,Xing_VR_Shannon,HABS+21,NGS22,mingzhe_TWC}. With these trajectory data, proactive streaming techniques based on the prediction of viewpoint become feasible~\cite{junnizou_TCSVT,yingcui_TIP,Xing_VR_Shannon}. These approaches proactively stream the content to be requested by users in advance~\cite{TRACK} and enable dynamic adaptation of video quality, such as tile quality adaptation \cite{yingcui_TIP}, according to time-varying viewpoint prediction errors and available resources. These proactive approaches significantly reduce the consumption of communication resources while effectively enhancing the quality of experience (QoE) of users~\cite{Xing_VR_Shannon,yingcui_TIP}.

However, every coin has two sides. During VR streaming, there is a risk that attackers might eavesdrop on these trajectory data \cite{UniqueIdentifNair,NGS22,WSZX+22} to deduce personal information \cite{WSZX+22,NGS22,MHJL+20}. Remarkably, with just a few minutes of seemingly-anonymous viewpoint and hand trajectories for VR video and VR games, 95\% of 511 users \cite{MHJL+20} and 94.33\% of over 50,000 users \cite{UniqueIdentifNair} can be identified.
Furthermore, a wide range of personal attributes, such as height, gender, age, and even more sensitive information like income, psychological attributes, sexual orientation, and women's hormonal cycles, can be inferred from these data~\cite{NGS22,privacy-preserving_eye_tracking_2021}. With the development of deep learning, the potential for attackers to extract more detailed personal information from VR trajectory data increases, raising serious concerns about trajectory leakage. Therefore, there is a growing demand from users for keeping trajectory data local~\cite{local_store,Privacy_VR_wx}, a requirement that aligns with legal regulations~\cite{gdpr}. This necessitates the development of privacy-preserving viewpoint prediction techniques that do not rely on uploading actual trajectory data, like federated training and local prediction~\cite{wx22,privacy_preserving_prediction,JSAC_private_VR}. However,
this approach only yields a fixed yet non-ideal level of privacy preservation~\cite{privacy_preserving_prediction}, which cannot satisfy the diverse privacy requirements.
This necessitates the incorporation of privacy-preserving data-processing approaches, which preserve privacy via reducing the usability of trajectory data \cite{nair22going}.

Existing privacy-preserving data-processing approaches for VR streaming mainly focused on gaze trajectory \cite{privacy-preserving_eye_tracking_2021,Differential_Privacy,privacy_def_eye_track,SHIM+19,
privacy_preserving_eye_dataset,BG21}, which can be classified into four categories:
(a) adding noise following Gaussian distribution~\cite{Differential_Privacy,privacy-preserving_eye_tracking_2021} or Laplace distribution~\cite{Differential_Privacy} to the actual trajectory data,
(b) temporal and spatial downsampling of the trajectory data~\cite{privacy-preserving_eye_tracking_2021,Differential_Privacy}, (c) combinations of downsampling and adding noise~\cite{privacy_def_eye_track,privacy_preserving_eye_dataset}, and
(d) learning-based approaches~\cite{privacy_preserving_eye_dataset}.
A few approaches considered viewpoint and hand trajectory~\cite{wx22,nair22going,privacy_preserving_prediction}, such as adding Laplace noise to the viewpoint and hand trajectory data~\cite{nair22going}, adding camouflaged tile requests around the actual field of view (FoV)~\cite{wx22}, or increasing the area of each tile~\cite{privacy_preserving_prediction}. The latter two approaches can be viewed as variants of adding noise to the viewpoint trajectory data. 
To evaluate the performance of these strategies, existing works focused on protecting specific personal information from being inferred using specific algorithms employed by attackers. For example, the inference of user identification by a radial basis function network is considered in \cite{privacy_preserving_eye_dataset}, and the inference of height and income of users by a transformed-based neural network is considered in \cite{nair2023inferring}. However, these evaluations using specific inference algorithms targeted at particular personal information may not adequately reflect the overall performance of information leakage. In reality, once attackers obtain the trajectory data, they could potentially utilize a range of sophisticated algorithms to infer any personal information that may have been overlooked during the preservation process. Therefore, the fundamental requirement for privacy preservation is to control the leakage of trajectory data itself. In the literature, however, the analysis of the leakage level of trajectory data has received limited attention.

Moreover, reducing the usability of trajectory data can have a negative impact on the performance of viewpoint prediction, ultimately leading to a degradation in the quality of experience (QoE) of the user. This creates a tradeoff between preserving privacy and enhancing QoE.
A few works evaluated the impact of existing data-processing approaches on prediction performance and privacy~\cite{privacy-preserving_eye_tracking_2021,Differential_Privacy,privacy_preserving_eye_dataset}. In~\cite{privacy-preserving_eye_tracking_2021}, it was shown that adding Gaussian noise increased the average gaze prediction errors by 1.15$^{\circ}$ and reduced the identification rate by 33\%.
In \cite{Differential_Privacy}, it was shown that adding Gaussian noise incurred less information loss of gaze heatmap than Laplace noise.
In~\cite{privacy_preserving_eye_dataset}, the Gaussian-based gaze sample replacement approach was employed, which increases the average gaze prediction errors from 6.8$^\circ$ to 9.1$^\circ$.
In summary, existing evaluations have shown the tradeoff between privacy and viewpoint prediction errors (directly causing QoE loss). However, the optimization of this tradeoff, particularly in minimizing the QoE loss while ensuring a specific required level of privacy, has not been addressed.

In this paper, we study trajectory leakage in the context of proactive VR streaming, where viewpoint prediction is involved. As demonstrated in \cite{TRACK}, the integration of gaze and hand trajectories alongside the viewpoint trajectory does not enhance the prediction performance of viewpoint when compared to relying solely on the viewpoint trajectory for prediction. Hence, we focus on the leakage of viewpoint trajectory data. We strive to analyze the probability of viewpoint leakage achieved by existing privacy-preserving approaches and propose a novel privacy-preserving approach, enabling the optimization of the tradeoff between QoE and privacy.

The main contributions are summarized as follows.
\begin{itemize}
\item We derive the probability of viewpoint leakage for existing approaches that preserve privacy by reducing the usability of actual viewpoint data. To this end, we determine the optimal viewpoint inferred by attackers and the optimal distribution of viewpoint prediction errors. The results show that existing approaches cannot fully eliminate the leakage of viewpoint information, which is mainly caused by the so-called ``PEA" relation, where the uploaded \emph{P}redicted viewpoint and prediction \emph{E}rrors are related to the \emph{A}ctual viewpoint. 
\item We propose a novel privacy-preserving approach that breaks the PEA relation by introducing noise to uploaded prediction errors. We demonstrate that the proposed approach is able to guarantee
zero leakage of viewpoint information. We then optimize the QoE-privacy tradeoff, aiming to minimize the loss of QoE while ensuring the requirement of viewpoint leakage probability. Considering that the distribution of prediction errors is unknown in practical applications, we conduct a worst-case optimization and find the optimal noise for any given value of prediction error. Simulation results demonstrate the advantages of the proposed approach in satisfying privacy requirement and reducing the loss of QoE compared to existing approaches.
\end{itemize}

The rest of the paper is organized as follows. In Sec. II, we introduce the system model. In Sec. III, we
analyze the probability of viewpoint leakage for existing privacy-preserving approaches. In Sec. IV, we propose a novel privacy-preserving approach and optimize the tradeoff between QoE and privacy. Sec. V presents simulation results and Sec. VI concludes the paper.

\begin{table*}
	\caption{Leaked information during viewpoint prediction}\label{table:FoV_leakage_one_manner}
	\begin{center}
		\begin{tabularx}{\textwidth}{
                        | >{\centering\arraybackslash}X
                        | >{\centering\arraybackslash}X
                        | >{\centering\arraybackslash}X
                        | >{\centering\arraybackslash}X
                        | >{\centering\arraybackslash}X
                        | >{\centering\arraybackslash}X
                        | >{\centering\arraybackslash}X
                        | >{\centering\arraybackslash}X
                        |}
			\hline
   			No.&Offline training mode$^{\dagger}$&Online predicting mode$^{\dagger}$&Data upload for offline training& Data upload in observation window& Data upload after online predicting& Refs.&Privacy preserving \\
                \cline{1-8}
			1&Centralized&Centralized&Real viewpoint& Real viewpoint& \backslashbox[1.1cm]{}{}&\cite{Xing_VR_Shannon} & \XSolid\\
			\cline{1-8}
			2&Centralized&Local&Real viewpoint& \backslashbox[1.1cm]{}{}&Predicted viewpoint&\cite{junnizou_TCSVT} & \XSolid \\
			\cline{1-8}
			\textbf{3}&Local&Local& \backslashbox[1.1cm]{}{}&\backslashbox[1.1cm]{}{}  & Predicted viewpoint&\cite{optimizing_VR} &\normalsize{?}\\
			\cline{1-8}
			\textbf{4}&Federated&Local&Model parameters$^{\ddagger}$ &\backslashbox[1.1cm]{}{}& Predicted viewpoint &\cite{JSAC_private_VR,wx22,privacy_preserving_prediction} &\normalsize{?}\\
			\cline{1-8}
			5 &Federated&Centralized&Model parameters & Real viewpoint&\backslashbox[1.1cm]{}{}&\backslashbox[0.5cm]{}{} & \XSolid \\
			\cline{1-8}
			6&No need&Centralized& \backslashbox[1.1cm]{}{} & Real viewpoint&\backslashbox[1.1cm]{}{} &\cite{TRACK} & \XSolid \\
			\cline{1-8}
			\textbf{7}&No need&Local& \backslashbox[1.1cm]{}{} & \backslashbox[1.1cm]{}{}&Predicted viewpoint&\cite{TRACK} &\normalsize{?}\\
			\hline
		\end{tabularx}
	\end{center}
	\footnotesize{ $^{\dagger}$ Centralized training or predicting is conducted at the server while local training and federated training are conducted at each user's HMD.}\\
        \footnotesize{$^{\ddagger}$ Model parameters of predictors, e.g., weights of a neural network.}
\end{table*}

\section{System Model}\label{section:system_model11}

\begin{figure}
    \centering
	\includegraphics[width=0.5\linewidth]{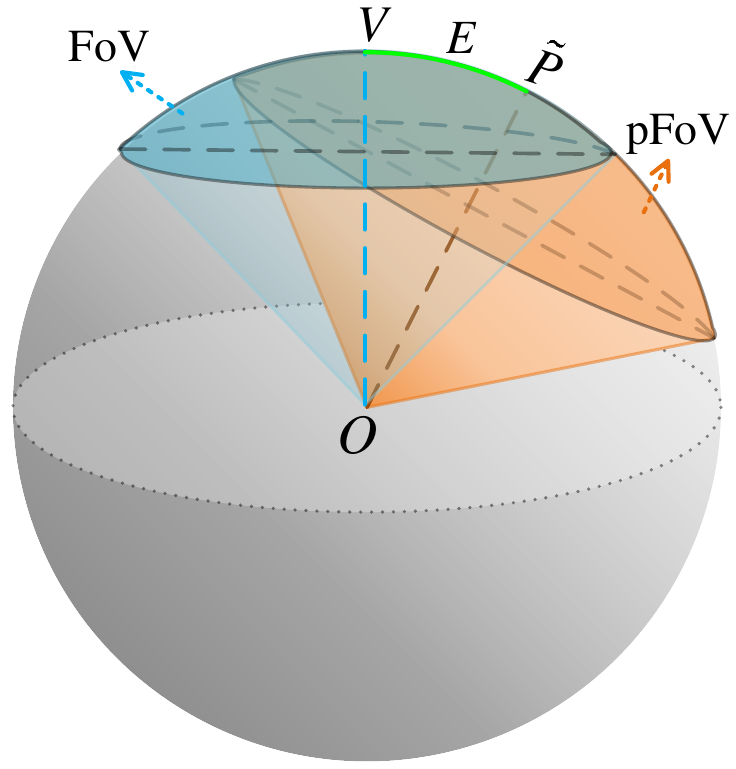}
    \captionsetup{font=small, labelsep=period}
    \caption{Illustration of FoV, pFoV, actual viewpoint $V$, predicted viewpoint $\widetilde{P}$, and prediction error $E$ on a unit sphere.}
    \label{Fig:fov}
\vspace{-0.3cm}
\end{figure}

Consider a proactive VR streaming system where users are served by an edge server co-located with a base station. The server either caches VR videos locally or accesses them through a high-speed wired connection, ensuring negligible delay from the Internet to the server. The users are allocated orthogonal resources for transmission, and thus we focus on the transmission and privacy concerns for a single user. Suppose that the user, positioned at the center of a unit sphere (denoted as $O$), watches a VR video through the FoV, as shown in Fig. \ref{Fig:fov}. The FoV is a spherical cap centered around the user's actual viewpoint (denoted as $V$).

Each VR video is spatially sliced into smaller units called tiles. These tiled videos are then encoded into different quality levels, each further divided into multiple groups of pictures (GoPs) with a uniform duration $T_{\mathrm{gop}}$  in the time dimension~\cite{VREXP}. 

\subsection{Viewpoint Uploading}
To support proactive VR streaming, the server requires the future viewpoints of users.\footnote{In this study, we focus on proactive VR streaming in the pushing mode, where the server receives uploaded viewpoints from all users and jointly optimizes which and how many tiles should be transmitted to each user. In contrast, in the pulling mode, each user typically uploads the tile indices that they wish to download, and the server has less control over the transmission compared to the pushing mode.} There have been several approaches in the literature to provide this information to the server. These approaches are summarized in Table~\ref{table:FoV_leakage_one_manner}, which highlights their differences in the mode of offline viewpoint predictor training and online predicting, as well as the uploaded data required for predictor training and predicting. The last column indicates whether an approach preserves user privacy or not.
We observe that approaches 3, 4, and 7 can prevent the leakage of the actual viewpoints during the training stage. Nevertheless, these approaches still require uploading the predicted viewpoint to the server, from which the actual viewpoint can be inferred~\cite{privacy_preserving_prediction,Xing_viewpoint_leakage_arxiv}.

To further preserve privacy, existing works have proposed to conduct proper data processing for the predictor. This involves modifying the input or output viewpoints of the predictor~\cite{Differential_Privacy,privacy-preserving_eye_tracking_2021,privacy_def_eye_track,privacy_preserving_eye_dataset,wx22,privacy_preserving_prediction,nair22going}.
For instance, in \cite{nair22going}, the actual viewpoint was modified by introducing Laplace noise, where the scale factor of the noise can be empirically adjusted to meet specific privacy requirements.

To reflect the current privacy-preserving design for viewpoint prediction, we consider that users do not upload their actual viewpoints. Instead, they upload the predicted viewpoint, which has undergone some form of data processing before being input to or output from the predictor.

\subsection{Prediction Errors Uploading}

\begin{figure}
	\centering
	\begin{minipage}[t]{1\linewidth}
		\includegraphics[width=1\textwidth]{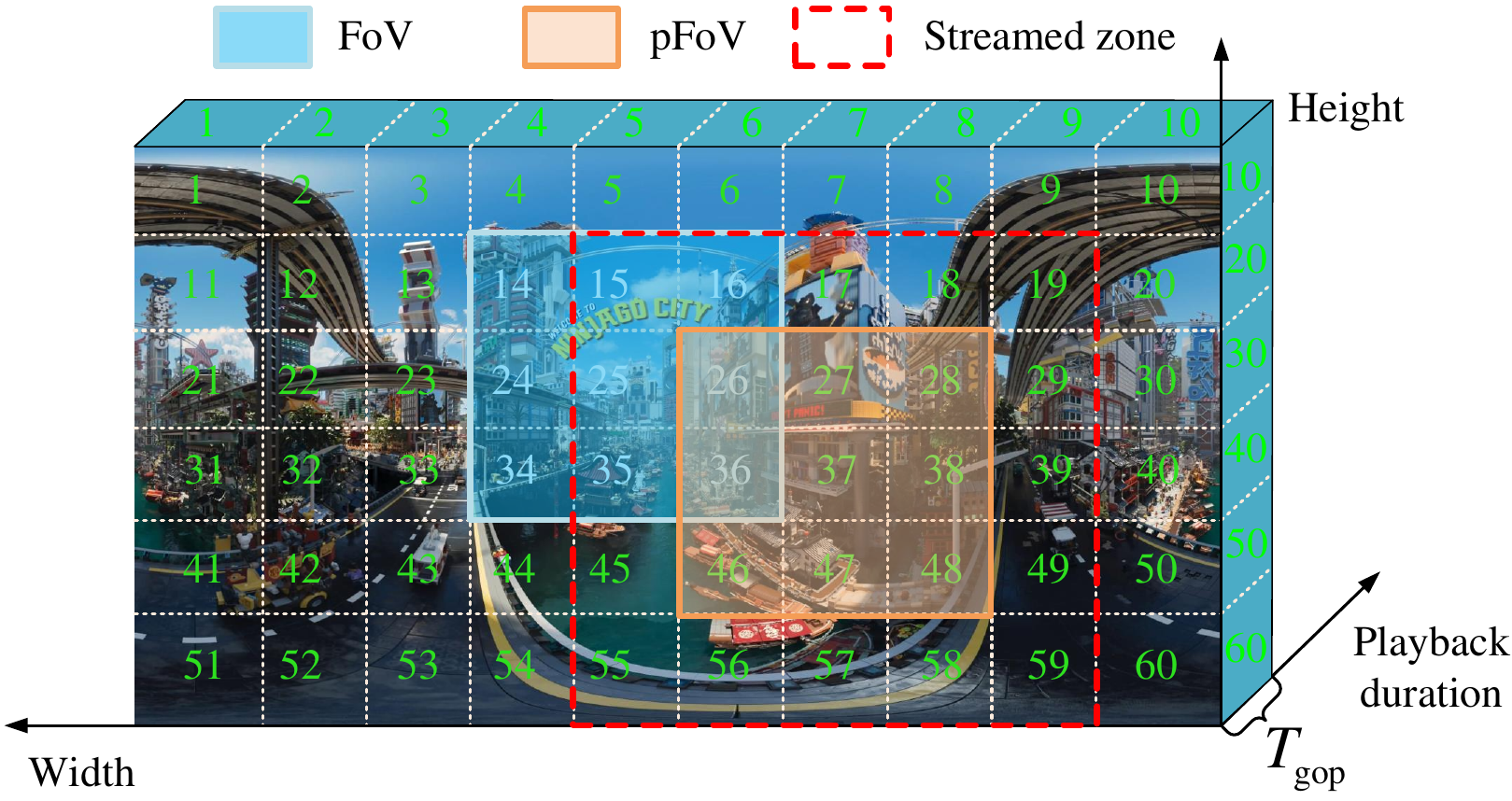}
	\end{minipage}
    \captionsetup{font=small, labelsep=period}
	\caption{Illustration of FoV, pFoV, and streamed zone in spatial-time~dimensions.}
	\label{Fig:FoV_2D}
	\vspace{-0.3cm}
\end{figure}

With the uploaded predicted viewpoint denoted as $\widetilde{P}$ from the user, the server can establish a predicted FoV (pFoV) as shown in Fig. \ref{Fig:fov}. By defining a streamed zone based on the pFoV, the server can proactively stream only the video tiles within this zone, enabling efficient content delivery~\cite{VREXP,yingcui_TIP}, as shown in Fig.~\ref{Fig:FoV_2D}.

However, due to the imperfect nature of viewpoint prediction, setting the streamed zone exactly as the pFoV will result in missing some tiles that are actually in the user's real FoV. Thus, it is necessary to make the streamed zone larger than the pFoV. To determine the appropriate size of the streamed zone, existing works have proposed allowing users to measure their viewpoint prediction errors and upload these error values to the server~\cite{changyangshe_VR_globecom,junnizou_TCSVT,metaverse_survey}.
The prediction error (denoted as $E$) is defined as the spherical distance between the predicted and actual viewpoints on a unit sphere~\cite{TRACK}, as shown in Fig. \ref{Fig:fov}. The prediction error $E$ has values within the range $[0,\pi]$ radians.

\begin{figure}
	\centering
	\begin{minipage}[t]{1\linewidth}
		\includegraphics[width=0.9\textwidth]{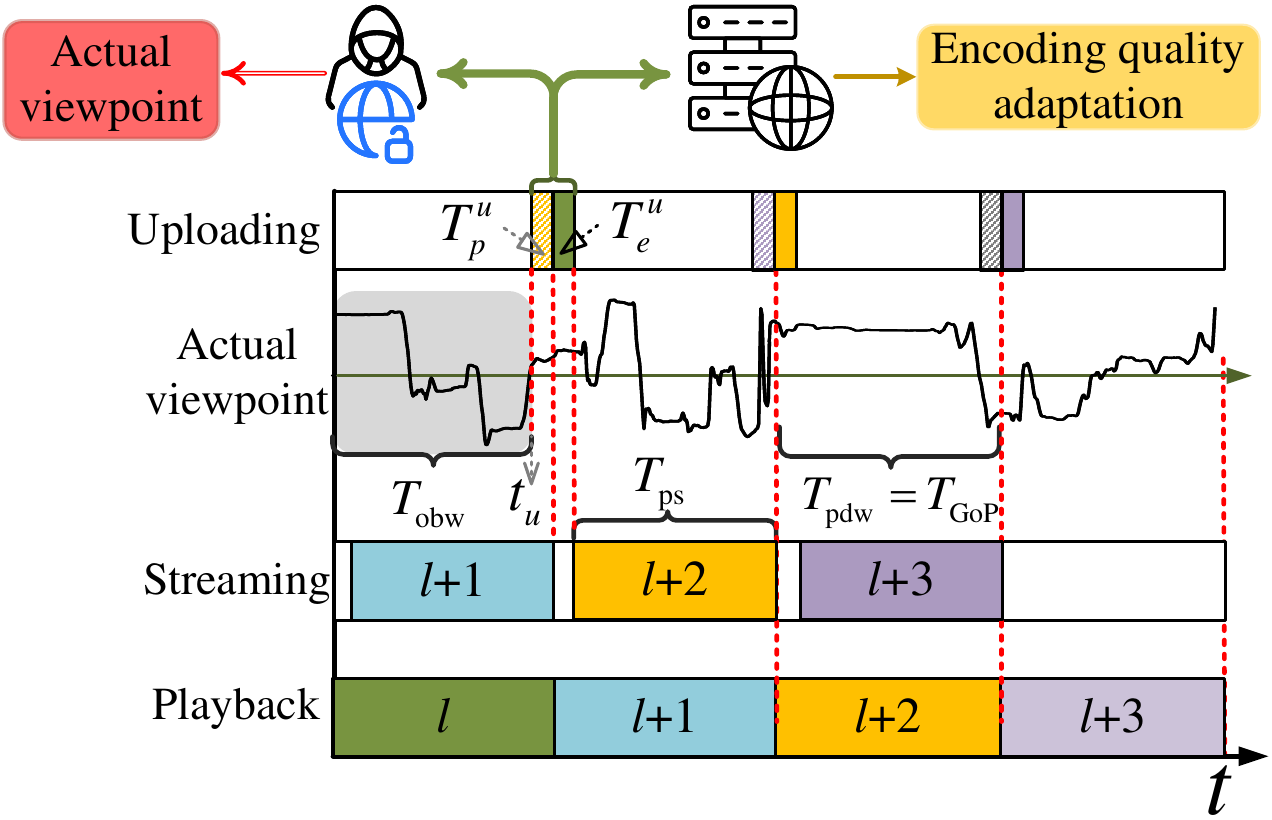}
	\end{minipage}
\captionsetup{font=small, labelsep=period}
	\caption{Proactive VR streaming with attackers}
	\label{Fig:proactive_streaming}
	\vspace{-0.3cm}
\end{figure}

Although uploading prediction errors can improve the performance of proactive VR streaming, it increases the risk of information leakage. We present an example process of proactive streaming in Fig.~\ref{Fig:proactive_streaming} to illustrate the uploaded and leaked information.
Let us consider the streaming of the ($l\!+\!2$)-th GoP.
After the playback of the $l$-th GoP, the head mounted displays (HMDs) of users compute the viewpoint prediction errors $E$ for the $l$-th GoP and upload the errors to the server within duration $T_{e}^u$. Prior to this, the predicted viewpoints $\widetilde{P}$ for the ($l\!+\!2$)-th GoP are uploaded within duration $T_{p}^u$, with which the pFoV is determined to estimate the tiles within the FoV. The server then uses the information about the tiles within the pFoV, prediction errors, and available communication resources to determine which tiles with what quality to stream for the ($l\!+\!2$)-th GoP \cite{Xing_VR_Shannon,soonbin_NOSSDAV}. These tiles are streamed within duration $T_{\mathrm{ps}}$.


\begin{figure}
\centering
    \includegraphics[width=0.5\linewidth]{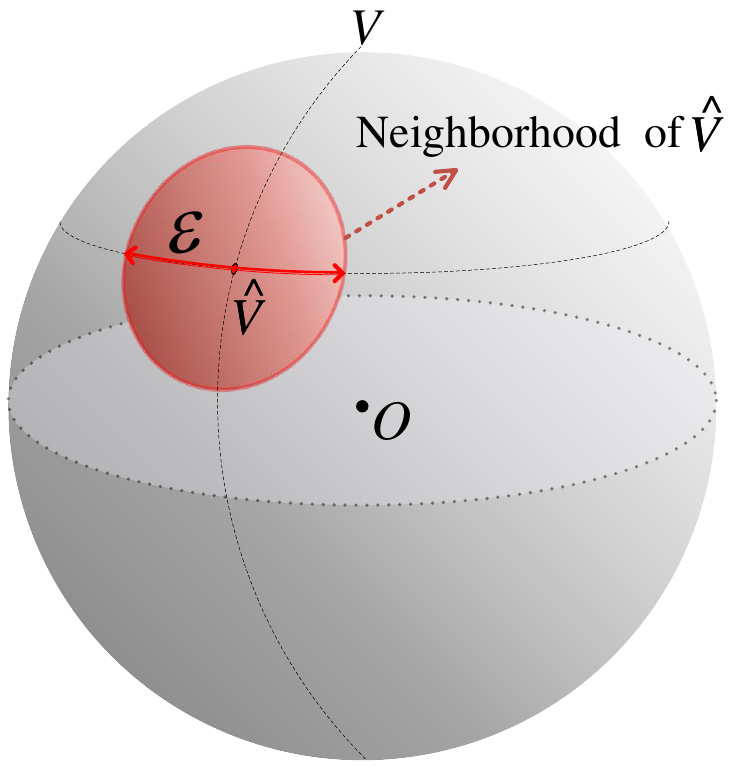}
    \captionsetup{font=small, labelsep=period}
    \caption{Neighborhood of $\widehat{V}$}
    \label{Fig:Neighbor_V}
\vspace{-0.3cm}
\end{figure}

\subsection{Privacy Attack}
In proactive VR streaming, the uploaded predicted viewpoints and prediction errors can be intercepted by attackers~\cite{UniqueIdentifNair,NGS22,WSZX+22}. Based on the intercepted information, attackers can infer the actual viewpoint with a certain level of precision (denoted as $\varepsilon$)
~\cite{privacy_def_eye_track,privacy_preserving_eye_dataset}, potentially revealing personal information about users, such as attributes and identities~\cite{MHJL+20,UniqueIdentifNair}.
Different personal information extraction tasks have different requirements on $\varepsilon$. For example, in the user identification scenario, a typical value for $\varepsilon$ is $0.1\pi$~\cite{privacy_preserving_eye_dataset}.

In Fig. \ref{Fig:Neighbor_V}, we illustrate the inferred viewpoint by $\widehat{V}$, whose neighborhood $\mathcal{N}(\widehat{V}, \varepsilon)$ is a spherical cap. If the spherical distance between the inferred and actual viewpoint does not exceed the required precision $\varepsilon$, the actual viewpoint falls within the neighborhood of the inferred viewpoint, i.e., $V\in\mathcal{N}(\widehat{V}, \varepsilon)$. We call this an event of viewpoint leakage, defined as $A:V\in\mathcal{N}(\widehat{V}, \varepsilon)$. Attackers aim to find the best inferred viewpoint $\widehat{V}$ that can maximize the viewpoint leakage probability $\mathrm{Pr}(A)$.

\subsection{QoE-Privacy Requirement}
For VR users, the privacy requirement can be expressed as $\mathrm{Pr}(A)\leq q$, where $q\in[0,1]$ is the user-specific privacy requirement. $q=0$ represents the most stringent privacy requirement, while $q=1$ indicates no privacy requirement. The conflict between privacy and QoE requirements is evident. Preserving privacy reduces the usability of the uploaded information for streaming~\cite{nair22going}, leading to a degradation in the QoE of VR streaming~\cite{Xing_viewpoint_leakage_arxiv}. We assume that as the accuracy of either the uploaded predicted viewpoints or prediction errors decreases, the QoE for the user also decreases. This assumption is supported by existing QoE models~\cite{Xing_VR_Shannon,VREXP,yingcui_TIP}.
In summary, the proactive streaming system aims to minimize the QoE loss while satisfying the privacy requirement $\mathrm{Pr}(A)\leq q$.

\section{Analysis of Viewpoint Leakage}\label{S:VL}
In this section, we analyze the viewpoint leakage probability $\mathrm{Pr}(A)$ and reveal the minimum leakage probability that existing privacy-preserving data processing approaches can~achieve.

\subsection{Viewpoint Leakage Probability} \label{S:s1}
To obtain $\mathrm{Pr}(A)$, we begin with deriving the conditional leakage probability $\mathrm{Pr}(A|\widetilde{P},e, \widehat{V})$ given the predicted viewpoint $\widetilde{P}$, prediction error $E=e$, and inferred viewpoint $\widehat{V}$.
We first find the optimal inferred viewpoint of attackers $\widehat{V}^*$ that maximizes the conditional probability, i.e.,
\begin{align}
  \widehat{V}^*=\arg\max_{\widehat{V}} \mathrm{Pr}(A|\widetilde{P},e, \widehat{V}),
\end{align}
with which we obtain the maximal conditional leakage probability.

\begin{figure}
	\centering
	\subfloat[$e\leq\varepsilon$]{\label{Fig:optimal_V_varepsilon_geq e}
		\begin{minipage}[c]{0.5\linewidth}
			\centering
			\includegraphics[width=1\textwidth]{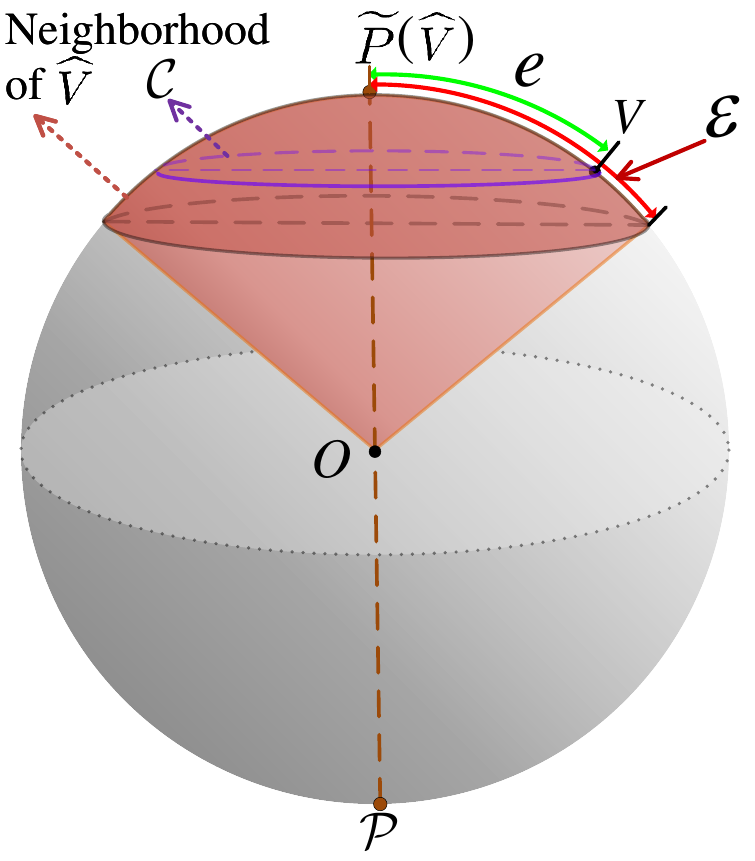}
		\end{minipage}
	}
	\subfloat[$e\geq \pi-\varepsilon$]{\label{Fig:optimal_V_varepsilon_geq pi_e}
		\begin{minipage}[c]{0.5\linewidth}
			\centering
			\includegraphics[width=1\textwidth]{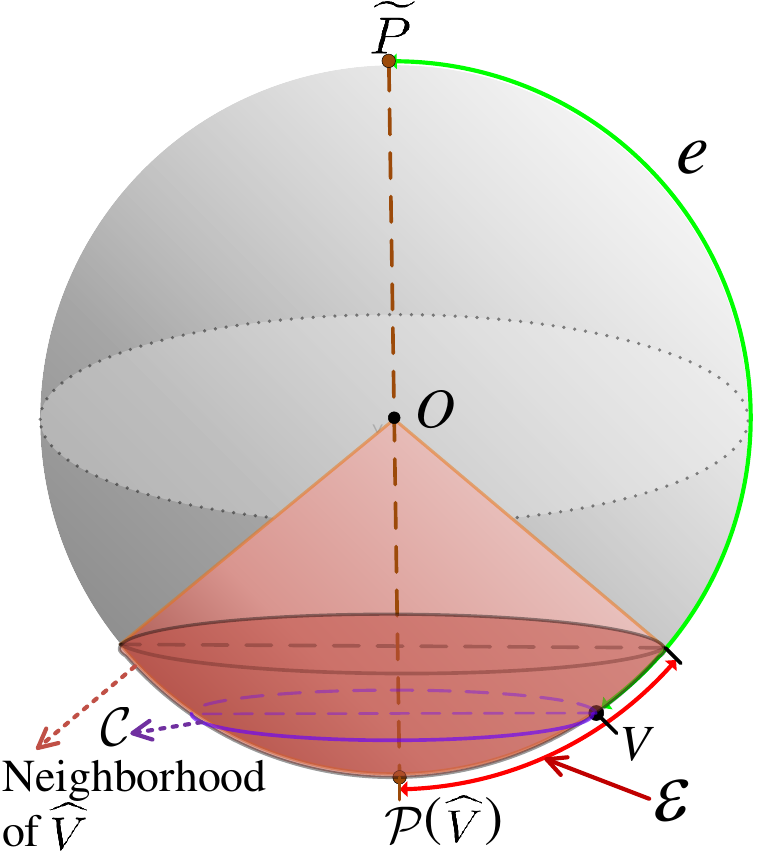}
		\end{minipage}
	}
\captionsetup{font=small, labelsep=period}
	\caption{Optimal $\widehat{V}$, $\mathrm{Pr}(A|\widetilde{P}, e, \widehat{V}^*)=1$. 
}\label{Fig:optimal_V}	
	\vspace{-0.3cm}
\end{figure}

\begin{figure}
	\centering
	\subfloat[$\pi\sin e>\varepsilon$]{\label{Fig:conditional_uncertainty_1}
		\begin{minipage}[c]{0.5\linewidth}
			\centering
			\includegraphics[width=1\textwidth]{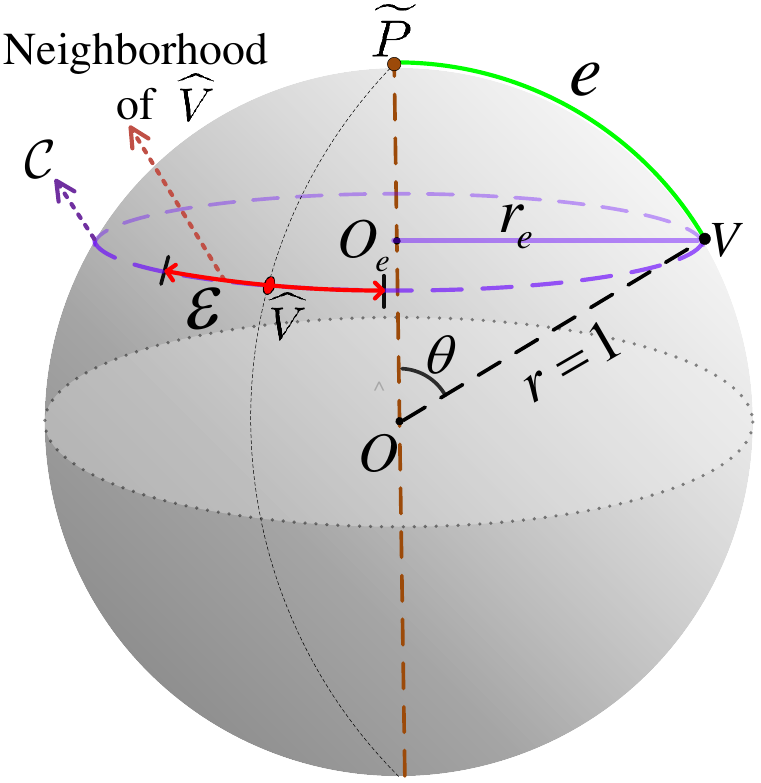}
		\end{minipage}
	}
	\subfloat[$\pi\sin e=\varepsilon$]{\label{Fig:conditional_uncertainty_2}
		\begin{minipage}[c]{0.5\linewidth}
			\centering
			\includegraphics[width=1\textwidth]{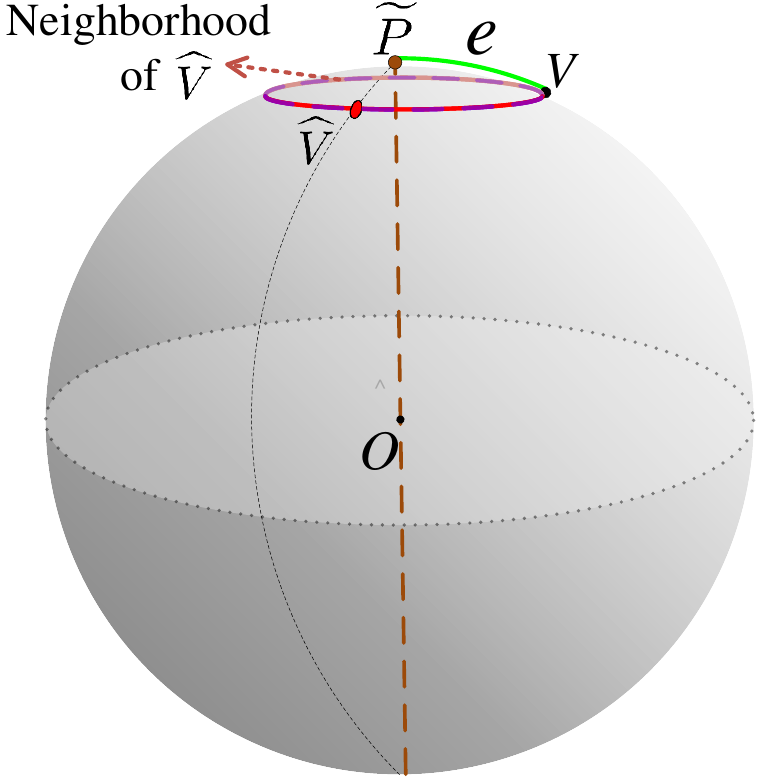}
		\end{minipage}
	}
\captionsetup{font=small, labelsep=period}
	\caption{$\mathrm{Pr}(A|\widetilde{P}, e)$ when $e\in[\varepsilon,\pi - \varepsilon]$.}\label{Fig:optimal_V_e_middle}	
	\vspace{-0.3cm}
\end{figure}

Prediction error measures the spherical distance between the predicted and actual viewpoints. Therefore, there exists a spherical relation among the Predicted viewpoint, prediction Error, and Actual viewpoint, which is referred to as the ``PEA" relation in the sequel. Based on the PEA relation,
as shown in Fig. \ref{Fig:optimal_V}, given the predicted viewpoint $\widetilde{P}$ and prediction error $e$, attackers can infer that the actual viewpoint is on a circle denoted by $\mathcal{C}$. According to the relationship between $e$ and $\varepsilon$, the inferred viewpoint $\widehat{V}^*$ can be determined in the following three cases.
\begin{itemize}
  \item Case 1: $e\leq \varepsilon$

  In this case, as shown in Fig. \ref{Fig:optimal_V_varepsilon_geq e}, attackers can select the predicted viewpoint $\widetilde{P}$ as the optimal inferred viewpoint $\widehat{V}^*$. Such a $\widehat{V}^*$ can ensure the occurrence of leakage event $A:V\in\mathcal{N}(\widehat{V}, \varepsilon)$, i.e., ensure that the actual viewpoint $V$ falls into the neighborhood of $\widehat{V}^*$.

  \item Case 2: $e\geq \pi - \varepsilon$

  In this case, as shown in Fig. \ref{Fig:optimal_V_varepsilon_geq pi_e}, attackers can select point $\mathcal{P}$ that is symmetric to the predicted viewpoint as the optimal inferred viewpoint $\widehat{V}^*$. It can be readily found that this $\widehat{V}^*$ can ensure the occurrence of leakage event $A$.

  \item Case 3: $e\in(\varepsilon,\pi - \varepsilon)$
  \begin{prop}\label{prop:optimal_v_circle_C}
    The optimal inferred viewpoint $\widehat{V}^*$ is a random point on the circle $\mathcal{C}$ where the actual viewpoint $V$ is located.

    Proof: See Appendix \ref{appendix:Prop_1}.
  \end{prop}

  Proposition 1 indicates that both the actual and inferred viewpoints are on the same circle $\mathcal{C}$. Then, the conditional leakage probability $\mathrm{Pr}(A|\widetilde{P},e, \widehat{V})$ equals to the probability that the spherical distance between $V$ and $\widehat{V}$ is not larger than $\varepsilon$.

  To obtain $\mathrm{Pr}(A|\widetilde{P},e, \widehat{V})$,
  we first determine the circumference of circle $\mathcal{C}$. As shown in Fig. \ref{Fig:optimal_V_e_middle}, the straight-line distance from any point on circle $\mathcal{C}$ to the center of the sphere $O$ is the sphere's radius $r=1$. The angle between the line $\overline{\widetilde{P} O}$ and the line $\overline{V O}$ is denoted as $\theta$, which equals to the spherical distance $e$ when $\theta$ is measured in radians. Therefore, the radius of circle $\mathcal{C}$ is obtained as $r_e = \sin\theta \cdot r = \sin e$. Consequently, the circumference of circle $\mathcal{C}$ is $2\pi \sin e$.

  When $2\varepsilon > 2\pi \sin e$, as shown in Fig. \ref{Fig:conditional_uncertainty_2}, the whole circle $\mathcal{C}$ falls within the neighborhood arc of $\widehat{V}$. In this situation, viewpoint leakage occurs with a probability of one. Otherwise, the probability of viewpoint leakage equals to the ratio of the length of the neighborhood arc to the circumference of circle $\mathcal{C}$, as shown in Fig. \ref{Fig:conditional_uncertainty_1}. Combining these two cases, we can obtain the conditional probability of viewpoint leakage for any given predicted viewpoint $\widetilde{P}$ and prediction error $e$ as
\begin{align}
 \!\!\! \mathrm{Pr}(A|\widetilde{P}, e, \widehat{V}) = \min\left(\frac{\varepsilon}{\pi \sin e},1\right), e\in(\varepsilon, \pi - \varepsilon).
  \label{lemma:Pr_main_case}
\end{align}
\end{itemize}

%
%
%

Summarizing the three cases, the maximal conditional leakage probability can be obtained~as
\begin{align}\label{E:I1_inst}
&\mathrm{Pr}(A|\widetilde{P}, e)  \triangleq  \max_{\widehat{V}} \mathrm{Pr}(A|\widetilde{P},e, \widehat{V}) \nonumber\\
&=\left
\{\begin{array}{lr}
1, &e\leq\varepsilon \ \textrm{or} \ e\geq\pi - \varepsilon,  \\
\min\left(\frac{\varepsilon}{\pi \sin e},1\right), &e\in(\varepsilon, \pi - \varepsilon).
\end{array}
\right.
\end{align}

When $\varepsilon \geq 0.5\pi$, we find from \eqref{E:I1_inst} that $(\varepsilon, \pi - \varepsilon)$ is empty and thus $\mathrm{Pr}(A|\widetilde{P}, e) = 1$ always holds, i.e., viewpoint leakage always happens. In practical scenarios, such an undesired result will not happen because usually the required precision $\varepsilon$ is much smaller than $0.5\pi$~\cite{nair22going,privacy_preserving_eye_dataset}. Therefore, in the sequel we only consider the scenario with $\varepsilon< 0.5\pi$.

By taking the expectation over $\widetilde{P}$ and $e$, the viewpoint leakage probability can be obtained as
\begin{align} \label{E:I1}
\mathrm{Pr}(A) &= \int_{e}\int_{\widetilde{P}} f_e(e) f_{\widetilde{P}}(\widetilde{P}) \mathrm{Pr}(A|\widetilde{P}, e) \mathrm{d} \widetilde{P} \mathrm{d} e \nonumber \\
&=\int_{0}^{\pi} f_e(e) \mathrm{Pr}(A|\widetilde{P}, e) \mathrm{d} e,
\end{align}
where $f_e(e)$ and $f_{\widetilde{P}}(\widetilde{P})$ denote the probability density functions (PDFs) of $\widetilde{P}$ and $e$, respectively, and the second equality holds because $\mathrm{Pr}(A|\widetilde{P}, e)$ is independent from $\widetilde{P}$ as shown by \eqref{E:I1_inst}.

Existing data-processing approaches preserve privacy by modifying the actual viewpoint data so as to increase prediction errors~\cite{privacy-preserving_eye_tracking_2021,Differential_Privacy,privacy_def_eye_track,SHIM+19,privacy_preserving_eye_dataset,
BG21,wx22,nair22going,privacy_preserving_prediction,JSAC_private_VR}.
Essentially, these approaches affect the PDF $f_e(e)$ of prediction errors.
This gives rise to two questions: What is the optimal  distribution of prediction errors for data-processing approaches to minimize viewpoint leakage probability?  Is it possible for the optimal distribution of prediction errors to entirely prevent viewpoint leakage? We address these questions in the next subsection.

\subsection{Distribution Optimization for Prediction Errors}
To find the optimal distribution of prediction errors, we formulate a variational problem to minimize the leakage probability as
\begin{subequations}\label{lemma:functional_prob}
\begin{align}
\min_{f_e(e)} &  \int_{0}^{\pi} f_e(e) \mathrm{Pr}(A|\widetilde{P}, e) \mathrm{d} e \label{lemma:functional_obj}\\
s.t. & \int_{0}^{\pi}f_e(e)\mathrm{d}e=1. \label{lemma:functional_cst}
\end{align}
\end{subequations}
The optimal solution to problem \eqref{lemma:functional_prob} is derived in Appendix \ref{appendix}, which can be expressed as
\begin{align}
  f_e^*(e)=\delta(e - 0.5\pi), \label{lemma:optimal_f_e}
\end{align}
where $\delta(\cdot)$ is the Dirac delta function.

The optimal distribution $f_e^*(e)$ indicates that the optimal data-processing approach that minimizes the viewpoint leakage probability should keep a fixed viewpoint prediction error at $0.5\pi$, which corresponds to the spherical distance covering half sphere. Such a large prediction error is clearly harmful to users' QoE, leading to a strong conflict between privacy-preserving and QoE enhancement.

However, even with such a large prediction error, by substituting \eqref{lemma:optimal_f_e} into \eqref{lemma:functional_obj}, we can obtain the minimal viewpoint leakage probability as
\begin{align}
    \mathrm{Pr}(A)_{\min}=\frac{\varepsilon}{\pi}>0. \label{lemma:min_mutual_information}
\end{align}

The result shows that even when the data-processing approaches are well designed to achieve the optimal distribution of prediction errors, the viewpoint leakage probability is still positive.
More importantly, existing data processing approaches cannot quantitatively control the distribution of prediction errors, and thus fail to guarantee the privacy requirement $\mathrm{Pr}(A)\leq q$. In the next section, we strive to develop an approach to ensure the privacy requirement.

\section{Optimizing QoE-Privacy Tradeoff}
As revealed by \eqref{E:I1}, the viewpoint leakage probability is affected by the distribution of prediction errors $f_e(e)$ and the conditional leakage probability $\mathrm{Pr}(A|\widetilde{P}, e)$. In the previous subsection, we have shown that existing data-processing approaches that affect $f_e(e)$ cannot guarantee the privacy requirement. In this section, we turn to control the conditional leakage probability $\mathrm{Pr}(A|\widetilde{P}, e)$, which is achieved by breaking the PEA relation
that connects the predicted viewpoint, prediction error, and actual viewpoint.

\subsection{Noisy Prediction Errors: Breaking PEA Relation}

We propose to break the PEA relation by adding noise to prediction errors. In this way, attackers can be misdirected to select the inferred viewpoint and then the conditional probability can be reduced.
We refer to the proposed approach that breaks the PEA relation as B-PEA.

In the following, we derive the conditional probability with noisy prediction errors, which is denoted as $\mathrm{Pr}(A|\widetilde{P}, e, n)\triangleq \max_{\widehat{V}}\mathrm{Pr}(A|\widetilde{P}, e, \widehat{V}, n)$, where $n$ denotes the added noise. 
Let $\widetilde{e}\triangleq e+n$ denote a noisy prediction error. If $\widetilde{e}<0$ or $\widetilde{e}>\pi$, attackers can infer that the uploaded prediction error is added with noise. To avoid this, we set $\widetilde{e}\in[0,\pi]$, and thus the value of $n$ ranges in $[-e,\pi-e]$.
Since attackers are unaware that the uploaded prediction error is added with noise, they take $\widetilde{e}$ as the actual prediction error to select the inferred viewpoint.



We next derive $\mathrm{Pr}(A|\widetilde{P}, e, n)$ in three cases.
\begin{itemize}
  \item Case A: $e\leq\varepsilon$

  Following the analysis for the three cases in Sec.~\ref{S:s1}, and considering that attackers select the optimal inferred viewpoint $\widehat{V}^*$ based on the noisy prediction error $\widetilde{e}$,
  we divide the analysis into three sub-cases according to the relationship between $\widetilde{e}$ and $\varepsilon$.
  \begin{itemize}
    \item[(i)] \underline{$\widetilde{e}\leq\varepsilon$ or equivalently $n\in[-e,\varepsilon-e]$}: This sub-case corresponds to Case 1 in Sec. \ref{S:s1}. Attackers select the predicted viewpoint $\widetilde{P}$ as the optimal inferred viewpoint $\widehat{V}^*$. Then, we can observe from Fig. \ref{Fig:optimal_V} that the actual viewpoint is still included in the neighborhood, and thus conditional probability with noise is $\mathrm{Pr}(A|\widetilde{P}, e, n)=1$.
\begin{figure}
	\vspace{-0.5cm}
	\centering
	\subfloat[$|n|<\varepsilon$]{\label{Fig:Conditional_entropy_BPEA_less_epsilon}
		\begin{minipage}[c]{0.48\linewidth}
			\centering
			\includegraphics[width=1\textwidth]{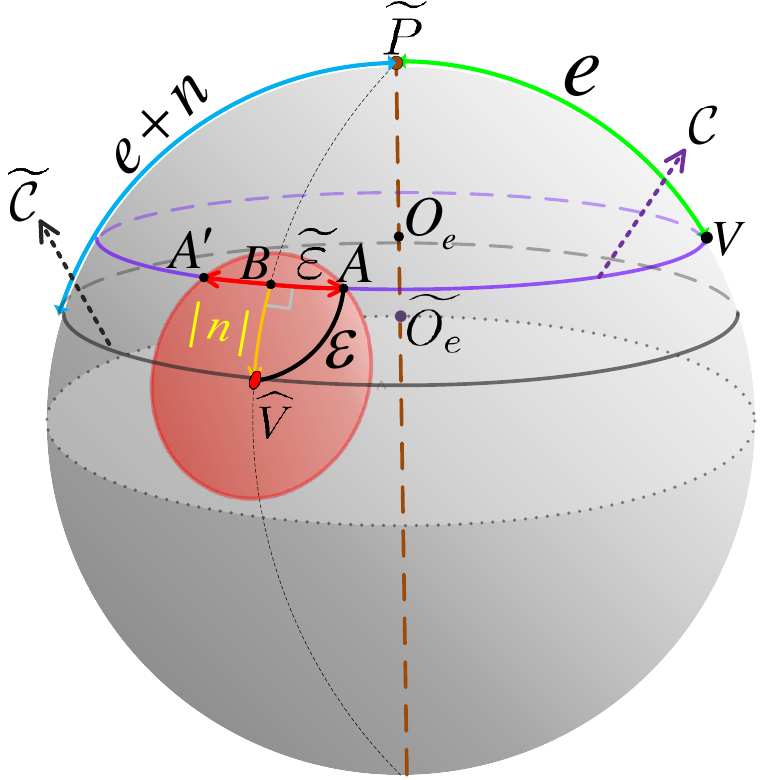}
		\end{minipage}
	}
	\subfloat[$|n|=\varepsilon$]{\label{Fig::Conditional_entropy_BPEA_equal_epsilon}
		\begin{minipage}[c]{0.48\linewidth}
			\centering
			\includegraphics[width=1\textwidth]{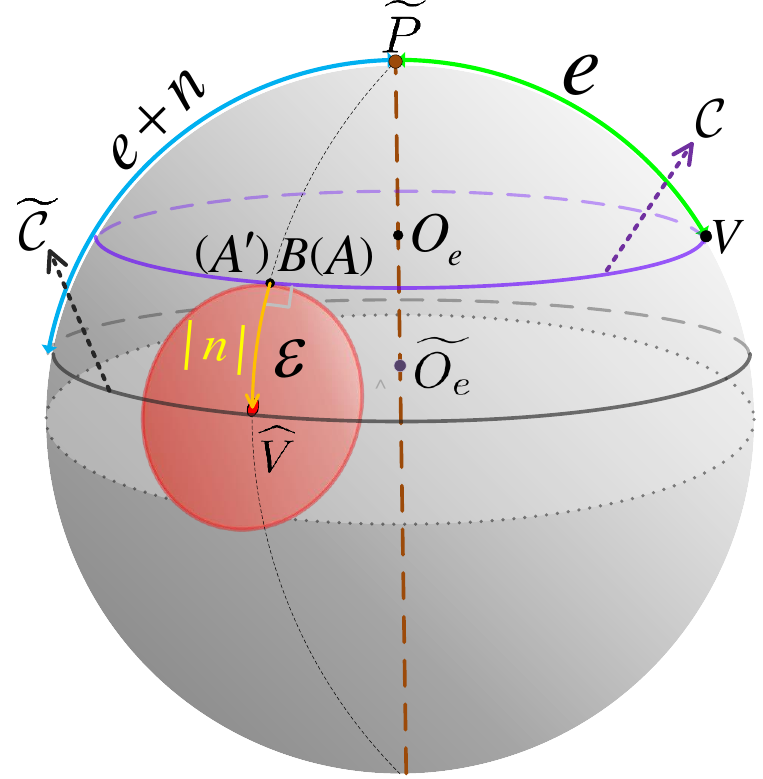}
		\end{minipage}
	}
\captionsetup{font=small, labelsep=period}
	\caption{$\mathrm{Pr}(A|\widetilde{P},e,n)$ when $n\in[\varepsilon - e,\pi - e - \varepsilon]$ }\label{Fig:Conditional_entropy_BPEA}	
	\vspace{-0.4cm}
\end{figure}

    \item[(ii)] \underline{$\widetilde{e}\!\in\![\varepsilon,\pi\!-\!\varepsilon]$ or equivalently $n\!\in\![\varepsilon \!-\! e,\pi \!-\! e \!-\! \varepsilon]$}: This sub-case corresponds to Case 3 in Sec.~\ref{S:s1}. Attackers select the predicted viewpoint $\widetilde{P}$ on a circle according to Proposition \ref{prop:optimal_v_circle_C}. Due to the existence of $n$, attackers infer that the actual viewpoint is on the circle $\widetilde{\mathcal{C}}$, as shown in Fig. \ref{Fig:Conditional_entropy_BPEA}.

        We next derive $\mathrm{Pr}(A|\widetilde{P}, e, n)$ for this sub-case. As shown in Fig. \ref{Fig:Conditional_entropy_BPEA}, let $B$ denote the point of intersection of the circle $\mathcal{C}$ and either the arc $\bigfrown{\widetilde{P}\widehat{V}}$  or its extension line, $\bigfrown{AA'}$ denote the arc of intersection of the circle $\mathcal{C}$ and the neighborhood of $\widehat{V}$, and $\widetilde{\varepsilon}$ denote a half of the arc length of $\bigfrown{AA'}$. The lengths of arc $\bigfrown{\widehat{V}B}$ and $\bigfrown{\widehat{V}A}$ are $|n|$ and $\varepsilon$, respectively. The circumference of the circle $\mathcal{C}$ is $2\pi\sin e$.

        Due to the introduction of $n$, the length of the neighborhood arc is reduced from $2\varepsilon$ in Fig. \ref{Fig:conditional_uncertainty_1} to $2\widetilde{\varepsilon}$ in Fig. \ref{Fig:Conditional_entropy_BPEA}.
        Then, by replacing $\varepsilon$ in \eqref{lemma:Pr_main_case} with $\widetilde{\varepsilon}$, we obtain $\mathrm{Pr}(A|\widetilde{P}, e, n)$ as
        \begin{align}
        \mathrm{Pr}(A|\widetilde{P}, e, n) = \min\left(\frac{\widetilde{\varepsilon}}{\pi \sin e}, 1\right),\label{lemma:Pr_BPEA_maincase_def}
        \end{align}
        where $\widetilde{\varepsilon}$ is derived as follows.

        When $|n|<\varepsilon$, as shown in Fig. \ref{Fig:Conditional_entropy_BPEA_less_epsilon}, Points $\widehat{V}$, $A$, and $B$ form a spherical right triangle. According to the spherical Pythagorean theorem, we have $\cos\varepsilon=\cos\widetilde{\varepsilon}\cos|n|$, from which we obtain $\widetilde{\varepsilon}=\arccos\left[\frac{\cos\varepsilon}{\cos|n|}\right]$.
        As the increase of $|n|$, the value of $\varepsilon$ decreases.
        When $|n|\geq\varepsilon$, as shown in Fig. \ref{Fig::Conditional_entropy_BPEA_equal_epsilon}, points $A$ and $A'$ converge to point $B$ and
        $\widetilde{\varepsilon}=0$.
        Therefore, $\widetilde{\varepsilon}$ can be expressed in a unified form as
        \begin{align}
        \widetilde{\varepsilon} = \arccos\left[\frac{\cos\varepsilon}{\cos\left(\min\left\{|n|,\varepsilon\right\}\right)}\right].
        \label{lemma:tilde_vareps}
        \end{align}
        Upon substituting \eqref{lemma:tilde_vareps} into \eqref{lemma:Pr_BPEA_maincase_def}, $\mathrm{Pr}(A|\widetilde{P}, e, n)$ is finally obtained as
        \begin{align}
          \mathrm{Pr}(A|\widetilde{P}, e, n) &\!=\!\min\!\left(\!\!\frac{\arccos\left[\frac{\cos\varepsilon}{\cos\left(\min\left\{|n|,\varepsilon\right\}\right)}\right]}{\pi \sin e}, 1\!\!\right).
          \label{lemma:Pr_BPEA_main_case}
        \end{align}
        We can observe that $\mathrm{Pr}(A|\widetilde{P}, e, n)$ is a decreasing function of $|n|$. When $|n|\geq\varepsilon$, $\mathrm{Pr}(A|\widetilde{P}, e, n)=0$.
        \item[(iii)] \underline{$\widetilde{e}\!\in\![\pi\!-\!\varepsilon,\pi]$ or equivalently $n\!\in\![\pi \!-\! e \!- \!\varepsilon, \pi \!-\! e]$}: This sub-case corresponds to Case 2 in Sec.~\ref{S:s1}. Attackers select the optimal inferred viewpoint $\widehat{V}^*$ as
            the symmetric point of the predicted viewpoint $\mathcal{P}$. As shown in Fig. \ref{Fig:BPEA_Pr_0_epsilon_large_than_e}, the neighborhood of the inferred viewpoint does not include the actual viewpoint, and thus $\mathrm{Pr}(A|\widetilde{P}, e, n)=0$.
  \end{itemize}

  \item Case B: {$e\!\in\!(\varepsilon,\pi-\varepsilon)$}

  The analysis can be divided into three sub-cases as Case A. For each sub-case attackers select the optimal inferred viewpoint $\widehat{V}^*$ as the same as Case A.
    \begin{itemize}
      \item[(i)] \underline{$\widetilde{e}\leq\varepsilon$ or equivalently $n\in[-e,\varepsilon - e]$}: Attackers select the optimal inferred viewpoint $\widehat{V}^*$ as the predicted viewpoint $\widetilde{P}$. As shown in Fig. \ref{Fig:Pr_BPEA_case2_i}, the neighborhood of the inferred viewpoint does not include the actual viewpoint, and thus $\mathrm{Pr}(A|\widetilde{P}, e, n)=0$.
      \item[(ii)] \underline{$\widetilde{e}\!\in\![\varepsilon,\pi\!-\!\varepsilon]$ or equivalently $n\!\in\![\varepsilon \!-\! e,\pi \!-\! e \!-\! \varepsilon]$}:
          Following the derivation for the sub-case A(ii), we can find that $\mathrm{Pr}(A|\widetilde{P}, e, n)$ in this sub-case is the same as that in the sub-case A(ii). The detailed derivation is omitted to avoid repetition.
      \item [(iii)] \underline{$\widetilde{e}\!\in\![\pi\!-\!\varepsilon,\pi]$ or equivalently $n\!\in\![\pi \!-\! e \!- \!\varepsilon, \pi \!-\! e]$}: Attackers select the optimal inferred viewpoint $\widehat{V}^*$ as the symmetric point of the predicted viewpoint $\mathcal{P}$. As shown in Fig. \ref{Fig:Pr_BPEA_case2_iii}, the actual viewpoint is not included in the neighborhood, and thus $\mathrm{Pr}(A|\widetilde{P}, e, n)=0$.
    \end{itemize}

\begin{figure}
	\vspace{-0.5cm}
	\centering
	\subfloat[$e\leq\varepsilon, \pi - e - \varepsilon\leq n\leq \pi - e$]{\label{Fig:BPEA_Pr_0_epsilon_large_than_e}
		\begin{minipage}[c]{0.42\linewidth}
			\centering
			\includegraphics[width=1\textwidth]{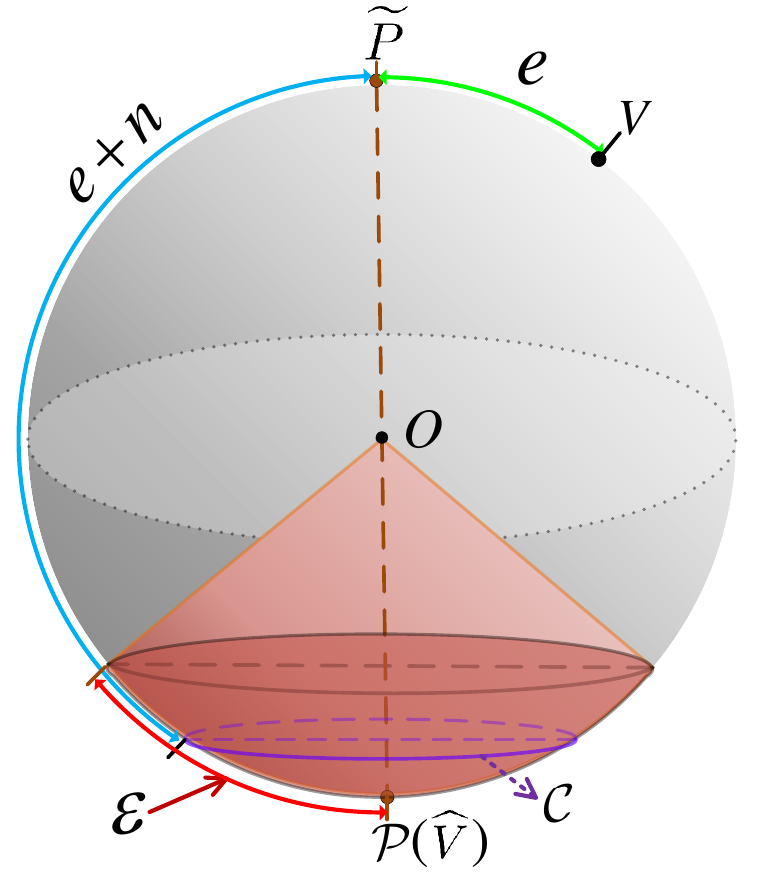}
		\end{minipage}
	}
	\subfloat[$e\!\in\!(\varepsilon,\pi-\varepsilon), n\in(-e,\varepsilon - e)$]{\label{Fig:Pr_BPEA_case2_i}
		\begin{minipage}[c]{0.42\linewidth}
			\centering
			\includegraphics[width=1\textwidth]{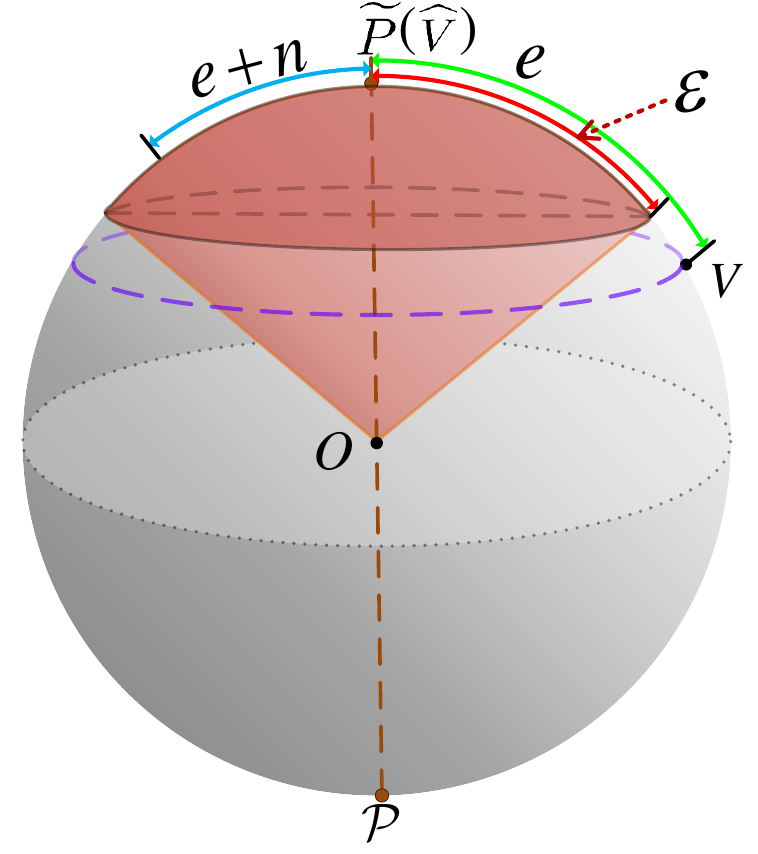}
		\end{minipage}
	}
\captionsetup{font=small, labelsep=period}
	\caption{$\mathrm{Pr}(A|\widetilde{P}, e, n)$ in sub-cases A(iii) and B(i). 
}\label{Fig:Pr_BPEA_case2}	
	\vspace{-0.4cm}
\end{figure}

  \item Case C: {$e\geq\pi-\varepsilon$}

  Again, the analysis is divided into three sub-cases as Case A and B.
  \begin{itemize}
    \item[(i)] \underline{$\widetilde{e}\leq\varepsilon$ or equivalently $n\in[-e,\varepsilon - e]$}: Attackers select the optimal inferred viewpoint $\widehat{V}^*$ as the predicted viewpoint $\widetilde{P}$. As shown in Fig. \ref{Fig:BPEA_Pr_0_epsilon_large_than_pi_e},
the neighborhood of the inferred viewpoint does not include the actual viewpoint, and thus $\mathrm{Pr}(A|\widetilde{P}, e, n)=0$.
    \item[(ii)] \underline{$\widetilde{e}\!\in\![\varepsilon,\pi\!-\!\varepsilon]$ or equivalently $n\!\in\![\varepsilon \!-\! e,\pi \!-\! e \!-\! \varepsilon]$}:
          Following the derivation for the sub-case A(ii), we can find that $\mathrm{Pr}(A|\widetilde{P}, e, n)$ in this sub-case is the same as that in the sub-case A(ii). The detailed derivation is omitted to avoid repetition.
    \item [(iii)] \underline{$\widetilde{e}\!\in\![\pi\!-\!\varepsilon,\pi]$ or equivalently $n\!\in\![\pi \!-\! e \!- \!\varepsilon, \pi \!-\! e]$}: Attackers select the optimal inferred viewpoint $\widehat{V}^*$ as the symmetric point of the predicted viewpoint $\mathcal{P}$. Then, we can observe from Fig. \ref{Fig:optimal_V} that the actual viewpoint is included in the neighborhood, and thus $\mathrm{Pr}(A|\widetilde{P}, e, n)=1$.
  \end{itemize}

\end{itemize}

 \begin{figure}
	\vspace{-0.5cm}
	\centering
	\subfloat[$e\!\in\!(\varepsilon,\pi-\varepsilon)$, $n\in(\varepsilon - e,\pi - e - \varepsilon)$]{\label{Fig:Pr_BPEA_case2_iii}
		\begin{minipage}[c]{0.46\linewidth}
			\centering
			\includegraphics[width=1\textwidth]{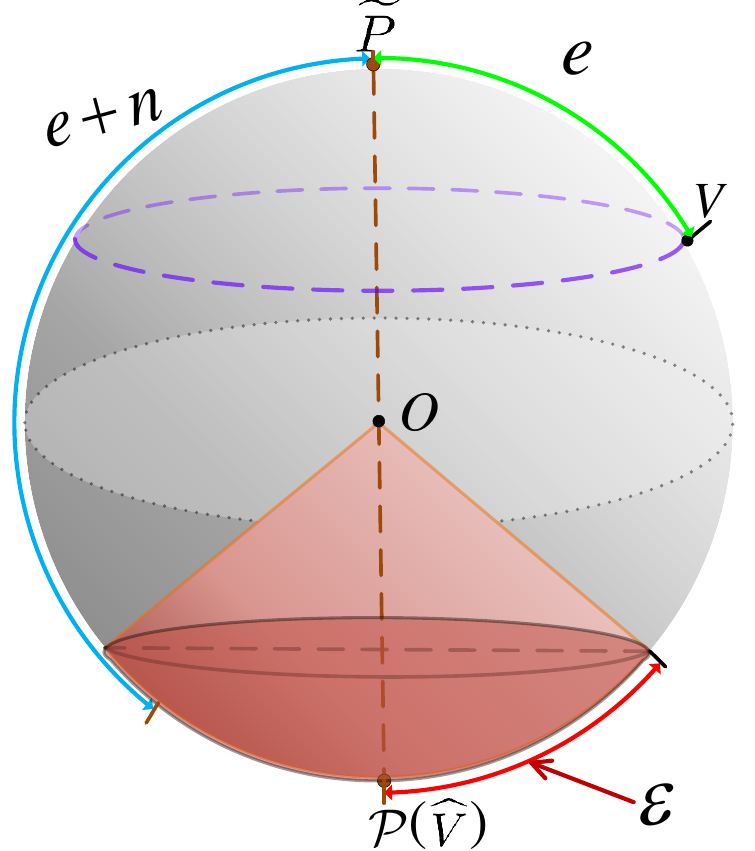}
		\end{minipage}
	}
	\subfloat[$e\geq \pi-\varepsilon$, $-e\leq n \leq\varepsilon - e$]{\label{Fig:BPEA_Pr_0_epsilon_large_than_pi_e}
		\begin{minipage}[c]{0.46\linewidth}
			\centering
			\includegraphics[width=1\textwidth]{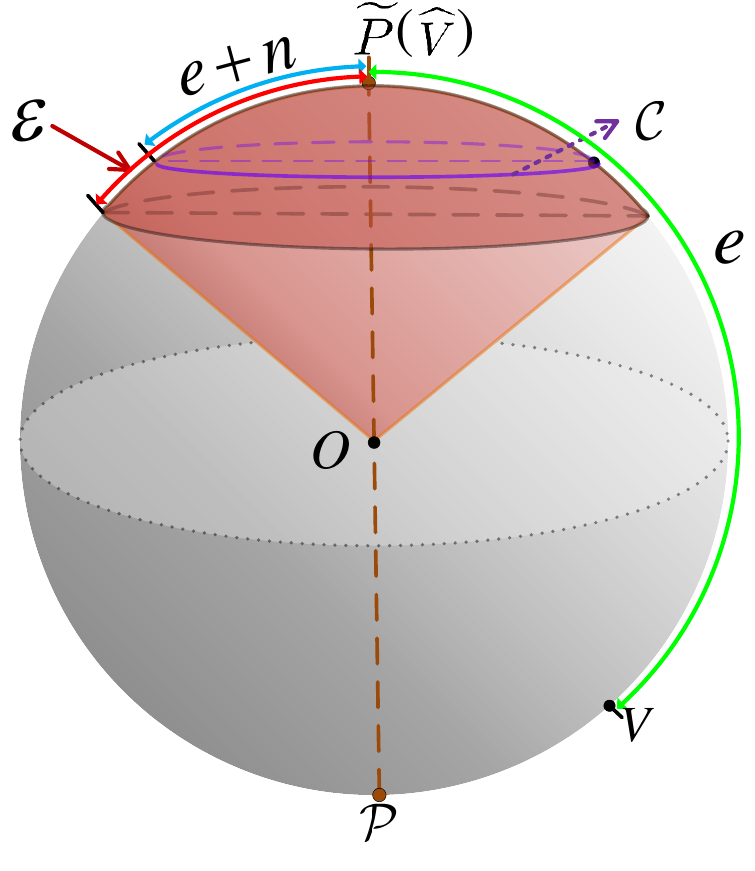}
		\end{minipage}
	}
\captionsetup{font=small, labelsep=period}
	\caption{$\mathrm{Pr}(A|\widetilde{P}, e, n)$ in sub-cases B(iii) and C(i). 
}\label{Fig:Pr_BPEA_case2}	
	\vspace{-0.4cm}
\end{figure}

Table \ref{table:I_e_eps_n} summarizes the conditional leakage probability $\mathrm{Pr}(A|\widetilde{P}, e, n)$ obtained in the above three cases, where the rows represent the three cases according to the value of $e$, and the columns represent the three sub-cases according to the value of $n$.
It can be found that for different values of $e$ (i.e., rows), we can always find a corresponding noise $n$ to make $\mathrm{Pr}(A|\widetilde{P}, e, n)$ be zero. It indicates that the proposed B-PEA has the capability of completely eliminating viewpoint leakage.

\begin{table}
\tabcolsep=0.005cm
	\captionsetup{font={small}}
	\caption{Summary of $\mathrm{Pr}(A|\widetilde{P}, e, n)$}\label{table:I_e_eps_n}
	\begin{center}
		\begin{tabular}{|c|c|c|c|}
			\hline
\diagbox{$e$}{$\mathrm{Pr}(A|\widetilde{P}, e, n)$}{$n$}&$n\!\in\![-e,\varepsilon - e]$&$n\!\in\!(\varepsilon \!-\! e,\pi \!-\! e \!-\! \varepsilon)$&$n\!\in\![\pi \!-\! e \!-\! \varepsilon, \pi \!-\! e]$\\
\hline
$e\leq \varepsilon$&1&&\multirow{2}{*}{0}\\
			\cline{1-2}
$e\!\in\!(\varepsilon,\pi-\varepsilon)$&\multirow{2}{*}{0}&Eq.\eqref{lemma:Pr_BPEA_main_case}&\\
\cline{1-1}\cline{4-4}
$e\geq\pi-\varepsilon$&&&1\\
\hline
		\end{tabular}
	\end{center}
\end{table}

Table \ref{table:I_e_eps_n} shows that a given value of noise $n$ can yield different values of $\mathrm{Pr}(A|\widetilde{P}, e, n)$, depending on the relation between $e$ and $\varepsilon$. This implies that $n$ is related to $e$ and $\varepsilon$. $n$ may be a deterministic function of $e$ and $\varepsilon$, or it may be a random variable related to $e$ and $\varepsilon$. Herein, we model $n$ as a random variable following a conditional PDF $f_{n|e,\varepsilon}(n)$. If $f_{n|e,\varepsilon}(n)$ is a Dirac delta function, $n$ reduces to a deterministic function of $e$ and $\varepsilon$.

In the next subsection, we optimize $f_{n|e,\varepsilon}(n)$ to minimize the impact of adding noise on QoE while satisfying the requirement of viewpoint leakage probability.

\subsection{Optimizing QoE-Privacy Tradeoff with B-PEA}

The proposed B-PEA introduces noise to the uploaded prediction error, which can mitigate the viewpoint leakage but at the same time will affect the determination of the streamed zone for proactive streaming. The latter degrades the QoE of users. To optimize the noise-adding strategy, we optimize the conditional PDF of noise $f_{n|e,\varepsilon}(n)$, aimed at minimizing the average absolute value of noise subject to the privacy requirement, which can be formulated~as
\begin{subequations}\label{lemma:functional_BPEA_prob}
\begin{align}
&\min_{f_{n|e,\varepsilon}(n)} \  \int_{e} f_e(e) \int_{n} |n| f_{n|e,\varepsilon}(n) \mathrm{d} n \mathrm{d} e \label{lemma:functional_BPEA_obj}\\
& s.t.\ \mathrm{Pr}(A) = \int_{e}f_e(e) \int_{n} f_{n|e,\varepsilon}(n) Pr(A|\widetilde{P}, e, n) \mathrm{d} n  \mathrm{d} e \leq q, \label{E:Pr_BPEA}\\
&\qquad \int_{n} f_{n|e,\varepsilon}(n) \mathrm{d} n = 1, \label{lemma:functional_BPEA_cst2}
\end{align}
\end{subequations}
where $Pr(A)$ is the probability of viewpoint leakage, and $q$ is the required maximal leakage probability.  Problem \eqref{lemma:functional_BPEA_prob} is a variational problem optimizing the conditional PDF $f_{n|e,\varepsilon}(n)$, whose solution is given by Proposition 2.

\begin{prop}\label{prop:optimal_f_n}
$f^{*}_{n|e,\varepsilon}(n)$ is a Dirac Delta function.\\
Proof: See Appendix \ref{appendix:optimal_f_e}.
\end{prop}

Proposition 2 indicates that $n$ is a deterministic function of $e, \varepsilon$, denoted by $n(e,\varepsilon)$. Then, problem \eqref{lemma:functional_BPEA_prob} can be transformed as
\begin{subequations}\label{lemma:optimal_n_e_eps_prob}
\begin{align}
\text{P0}: \min_{n(e,\varepsilon)} &\ \int_{e} |n(e,\varepsilon)|f_e(e) \mathrm{d} e \label{lemma:optimal_n_e_eps_prob_obj}\\
s.t. &\ \mathrm{Pr}(A) =\int_{e}f_e(e) Pr\left(A|\widetilde{P}, e, n(e,\varepsilon)\right) \mathrm{d} e\leq q. \label{lemma:optimal_n_e_eps_prob_cst}
\end{align}
\end{subequations}

A difficulty of solving problem P0 lies in that the distribution of prediction error $e$ is unknown during the streaming. To deal with this issue, we conduct a worst-case study by obtaining an upper bound of $\mathrm{Pr}(A)$, which does not depend on the distribution of $e$.
Specifically, noting that $\int_{e}f_e(e) Pr\left(A|\widetilde{P}, e, n(e,\varepsilon)\right) \mathrm{d} e\leq \max_{e} Pr\left(A|\widetilde{P}, e, n(e,\varepsilon)\right) $, thus problem P0 can be tightened as
\begin{subequations}\label{lemma:optimal_n_e_eps_prob}
\begin{align}
\text{P1}: \min_{n(e,\varepsilon)} &\ \int_{e} |n(e,\varepsilon)|f_e(e) \mathrm{d} e \label{lemma:optimal_n_e_eps_prob_obj}\\
s.t. &\ \max_{e} Pr\left(A|\widetilde{P}, e, n(e,\varepsilon)\right) \leq q. \label{lemma:optimal_n_e_eps_prob_cst}
\end{align}
\end{subequations}
Problem P1 is a tightened version of P0 because the feasible region of P1 is smaller than that of P0.

The constraint on the maximal $Pr\left(A|\widetilde{P}, e, n(e,\varepsilon)\right)$ for all possible $e$ in \eqref{lemma:optimal_n_e_eps_prob_cst} can be equivalently transformed into the constraint for every $e$ as
\begin{align}
  Pr\left(A|\widetilde{P}, e, n(e,\varepsilon)\right) \leq q, \ \forall e.
\end{align}
With the constraint for every $e$, the objective function taking the expectation over $e$ can be decoupled for every $e$. This leads to the following problem for every given $e$:
\begin{subequations}\label{lemma:optimal_n_e_eps_prob_3}
\begin{align}
\text{P2}: \min_{n(e,\varepsilon)}& \ |n(e,\varepsilon)| \\
s.t. \ &Pr\left(A|\widetilde{P}, e, n(e,\varepsilon)\right)\leq q.
\label{c:BPEA_uncertainty_inst}
\end{align}
\end{subequations}

Based on the result of $Pr\left(A|\widetilde{P}, e, n(e,\varepsilon)\right)$ given in Table \ref{table:I_e_eps_n}, we next solve problem P2 in three cases according to the value of $e$, respectively. For each case, we further consider three sub-cases according to the value of $n$. We first find the optimal solution for each sub-case, and then obtain the optimal solution by comparing these solutions. For notational simplicity, we denote $Pr\left(A|\widetilde{P}, e, n(e,\varepsilon)\right)$ by $p(n)$ in the~sequel.

\begin{itemize}
  \item Case A: $e\leq \varepsilon$
  \begin{itemize}
    \item[(i)] \underline{$n\!\in\![-e,\varepsilon - e]$}: As shown in Table~\ref{table:I_e_eps_n}, $p(n)$ equals 1 in this sub-case, which cannot satisfy the constraint \eqref{c:BPEA_uncertainty_inst}.
    \item[(ii)] \underline{$n\!\in\!(\varepsilon - e,\pi - e - \varepsilon)$ }: In this sub-case $n$ is non-negative since $e\leq \varepsilon$ in Case A, and $p(n)$ is given by \eqref{lemma:Pr_BPEA_main_case}, which is a decreasing function of $n$. As a result, if $p(\varepsilon-e) < q$, the optimal $n$ that minimizes $|n|$ is $n^*=\varepsilon-e+\tau$, where $\tau>0$ is an arbitrarily small constant to ensure that $n>\varepsilon - e$. If $p(\pi - e -\varepsilon) >q$, there is no feasible solution. If $p(\varepsilon-e) < q<p(\pi - e -\varepsilon)$, by substituting \eqref{lemma:Pr_BPEA_main_case} into \eqref{c:BPEA_uncertainty_inst}, after some regular derivations, we have
        \begin{align}
        \!\!n^*\!=\! \mathscr{N}(e,\varepsilon,q)\triangleq\arccos\left( \frac{\cos\varepsilon}{\cos\left(q\pi\sin e\right)} \right),
        \end{align}
        where $\mathscr{N}(e,\varepsilon,q)\in(\varepsilon - e,\pi - e - \varepsilon)$.

    \item[(iii)] \underline{$n\!\in\![\pi - e - \varepsilon, \pi - e]$}: $p(n)$ equals 0 in this sub-case, which can always satisfy the constraint. Since we consider $\varepsilon < 0.5\pi$ and $e\leq \varepsilon$ in Case A, we have $\pi - e - \varepsilon >0$. Therefore, the optimal $n$ with the minimal $|n|$ is $n^* = \pi - e - \varepsilon$.
  \end{itemize}

  Comparing the optimal solutions of the three sub-cases, we obtain $n^*$ for Case A as
    \begin{align}
    n^*=
    \left
    \{\begin{array}{ll}
    \varepsilon-e+\tau, &  p(\varepsilon-e)\leq q,    \\
    \pi - e - \varepsilon, &  p(\pi - e - \varepsilon)\geq q, \\
    \mathscr{N}(e,\varepsilon,q), &\ \text{otherwise}.
    \end{array}
    \right.
    \end{align}

    \item Case B: $e\in(\varepsilon, \pi-\varepsilon)$

    \begin{itemize}
    \item[(i)] \underline{$n\!\in\![-e,\varepsilon - e]$}: $p(n)$ equals 0 in this sub-case, which always satisfies the constraint \eqref{c:BPEA_uncertainty_inst}. Since $e\in(\varepsilon, \pi-\varepsilon)$ in Case B, we have $\varepsilon - e<0$. Thus, the optimal $n$ is  $n^*= \varepsilon - e$.

    \item[(ii)] \underline{$n\!\in\!(\varepsilon - e,\pi - e - \varepsilon)$ }: In this sub-case, the lower and upper limits of $n$ is negative and positive, respectively, and $p(n)$ is given by \eqref{lemma:Pr_BPEA_main_case}, which is a decreasing function of $|n|$. As a result, if $p(\max{\{|\varepsilon-e|, \pi - e - \varepsilon\}}) >q$, then there is no feasible solution. If $p(0) \leq q$, the optimal solution is $n^*=0$. If $p(0) < q<p(\max{\{|\varepsilon-e|, \pi - e - \varepsilon\}})$, similar to sub-case A(ii), we can obtain $|n^*|= \mathscr{N}(e,\varepsilon,q)$, where $\mathscr{N}(e,\varepsilon,q)\leq \max{\{|\varepsilon-e|, \pi - e - \varepsilon\}}$.

    \item[(iii)] \underline{$n\!\in\![\pi - e - \varepsilon, \pi - e]$}: $p(n)$ equals 0 in this sub-case, which always satisfies the constraint \eqref{c:BPEA_uncertainty_inst}. Note that the lower limit of $n$ is positive, thus the optimal $n$ with minimal $|n|$ is $\pi - e - \varepsilon$.
  \end{itemize}

  We obtain the optimal value of $n^*$ by comparing the three sub-cases, which is summarized as follows.
  \begin{itemize}
     \item If $p(0)\leq q$, $n^*=0$, which is achieved in sub-case(ii).
     \item If $p(\max\{|\varepsilon-e|, \pi\!-\!e\!-\!\varepsilon\})\geq q$, sub-case(ii) is infeasible and the optimal $n^*$ is
            \begin{align}
              n^* = \left\{
              \begin{array}{ll}
                \varepsilon - e,   & \ |\varepsilon-e|\leq\pi \!-\!e \!-\! \varepsilon,  \\
                \pi \!-\!e \!-\! \varepsilon, & \ \text{otherwise}.
                \end{array}
                \right.
            \end{align}
     \item If $p(\max\{|\varepsilon-e|, \pi\!-\!e\!-\!\varepsilon\})< q < p(0)$, the three sub-cases are all feasible. Defining $\eta \triangleq \min\{|\varepsilon-e|,\pi \!-\! e \!-\! \varepsilon, \mathscr{N}(e,\varepsilon,q)\}$, then the optimal $n^*$ is
            \begin{align}
              \!\!\!n^* = \left\{
              \begin{array}{ll}
                \varepsilon - e,   & \ |\varepsilon-e|=\eta,  \\
                \pi \!-\!e \!-\! \varepsilon, & \ \pi \!-\!e \!-\! \varepsilon=\eta, \\
                \pm\mathscr{N}(e,\varepsilon,q), & \ \mathscr{N}(e,\varepsilon,q)\varepsilon=\eta.  \end{array}
                \right.
            \end{align}
  \end{itemize}

    \item Case C: $e\geq\pi-\varepsilon$
    \begin{itemize}
    \item[(i)] \underline{$n\!\in\![-e,\varepsilon - e]$}: As shown in Table~\ref{table:I_e_eps_n}, $p(n)$ equals 0 in this sub-case, which can always satisfy the constraint \eqref{c:BPEA_uncertainty_inst}. Since $e\geq\pi-\varepsilon$ in Case C and $\varepsilon < 0.5\pi$, we have $\varepsilon -e <0$. Therefore, the optimal $n$ with the minimal $|n|$ is $n^* = \varepsilon -e$.

    \item[(ii)] \underline{$n\!\in\!(\varepsilon - e,\pi - e - \varepsilon)$ }: In this sub-case $n$ is negative since $\pi - e - \varepsilon<0$ in Case C, and $p(n)$ is given by \eqref{lemma:Pr_BPEA_main_case}, which is an increasing function of $n$. As a result, if $p(\pi - e - \varepsilon) <q$, then the optimal $n$ that minimizes $|n|$ is $n^*=\pi - e - \varepsilon-\tau$, where $\tau>0$ is to ensure that $n<\pi - e - \varepsilon$. If $p(\varepsilon - e) >q$, there is no feasible solution. If $p(\varepsilon - e) < q< p(\pi - e - \varepsilon)$, we can obtain that the optimal $n^*$ is $-\mathscr{N}(e,\varepsilon,q)$.

    \item[(iii)] \underline{$n\!\in\![\pi - e - \varepsilon, \pi - e]$}: $p(n)$ equals 1 in this sub-case, which cannot satisfy the constraint \eqref{c:BPEA_uncertainty_inst}.

  \end{itemize}

  Comparing the optimal solutions of the three sub-cases, we obtain $n^*$ for Case C as
    \begin{align}
    n^*=
    \left
    \{\begin{array}{ll}
\pi -e - \varepsilon-\tau, &  p(\pi -e - \varepsilon)\leq q,  \\
\varepsilon - e, &  p(\varepsilon-e)\geq q, \\
-\mathscr{N}(e,\varepsilon,q), & \ \text{otherwise}.
\end{array}
    \right.
    \end{align}

\end{itemize}

In practical applications, the user compares the measured prediction error $e$ with the required viewpoint inference precision $\varepsilon$ to determine which of the three cases applies, and then selects the corresponding optimal noise added to $e$ before uploading it.

\section{Trace-Driven Simulation Results}

In this section, we evaluate the performance of B-PEA based on a practical VR streaming platform and a state-of-the-art (SOTA) viewpoint predictor.

We consider the viewpoint prediction on a viewpoint trajectory dataset \cite{dataset_head},
which contains the unit quaternion traces from $K\!\!=\!\!48$ users watching nine VR videos.
These traces are first downsampled to five samples per second, and then transformed into three-dimensional coordinates. Each of these values is within [-1,1] \cite{TRACK}. Then, the pre-processed traces of $N_{\mathrm{train}}\!\!=\!\!5$ videos and $N_{\mathrm{test}}\!\!=\!\!4$ videos are respectively used as training and testing sets \cite{TRACK}.

According to the evaluation in \cite{TRACK}, a predictor named ``deep-position-only'' achieves SOTA prediction performance on the dataset when the length of the prediction window is less than three seconds.
The predictor employs a long-short-term-memory neural network, which uses the three-dimensional coordinates of viewpoint trajectory series in an observation window to predict the coordinates of viewpoint sequence in a prediction window \cite{TRACK}. In the default setting of the predictors, the prediction window starts immediately after the observation window. To reserve time for proactively streaming tiles, uploading predicted viewpoints, and uploading prediction errors before playback, we tailor the predictor by setting the duration between the end of the observation window and the beginning of the prediction windows as one second. Within the interval, the durations for uploading predicted viewpoints and errors as well as proactively streaming tiles are respectively $T_{p}^u=0.05$ s, $T_{e}^u =0.05$ s, and $T_{\mathrm{ps}}=0.95$ s. The durations for observation window and prediction window are set to $T_{\mathrm{obw}}=T_{\mathrm{pdw}}=1$ s \cite{TRACK,VREXP}.
For the privacy-preserving viewpoint prediction, we consider federated training and local predicting at each HMD. When training the predictor, we employ a classical federated learning algorithm known as \texttt{FederatedAveraging}~\cite{google_federated_learning_17}. The settings of the federated learning are as follows. In each round, the model parameters of the predictor are updated from the HMDs of all users. The number of local epochs for each user is $E_l=50$, and the number of communication rounds is $R=10$.
The weighting coefficient of the $k$th user on the model parameter is $c_k =\frac{1}{48}$. Other details and hyper-parameters of this predictor are the same as the deep-position-only predictor \cite{TRACK}.
After the online prediction with the well-trained predictor, the prediction errors for all viewpoint samples in the testing set can be obtained and subsequently averaged.

We evaluate the viewpoint leakage probability using the sample mean for existing data-processing approaches, which can be expressed as
\begin{align*}
\mathrm{Pr}(A)\approx&\frac{1}{N}\sum_{i=1}^N\Big(\mathbbm{1}(e_i\!\!\in\!\![\varepsilon,{\pi-\varepsilon}]) \min\left(\frac{\varepsilon}{\pi \sin e},1\right) + \\ &\mathbbm{1}(e_i\!\notin\![\varepsilon,{\pi-\varepsilon}])\Big)
\end{align*}
where $\mathbbm{1}(x)=1$ if $x$ is true and $0$ otherwise, and $N$ is the number of samples in the whole testing set or a trace of the testing set. For the proposed B-PEA,
$\mathrm{Pr}(A)\approx \frac{1}{N} \sum_{i=1}^N Pr\left(A|\widetilde{P}, e_i, n^*(e_i,\varepsilon)\right)$, where $n^*(e_i,\varepsilon)$ is the solution of problem P2. The parameter $\tau$ when computing  $n^*(e_i,\varepsilon)$ is set as $\tau=10^{-4}$. The required viewpoint inference precision $\varepsilon$ is set as $\varepsilon=0.1\pi$~\cite{privacy_preserving_eye_dataset}.

We realize the proactive VR streaming based on VR-EXP \cite{VREXP}, an open-source VR video streaming platform. The platform provides proactive tile-based VR streaming in multiple quality representations.
The videos are encoded into 4$\times$8 tiles with three quality representations i.e., 720p (1.8 Mbps), 1080p (2.7 Mbps), and 4K (6 Mbps), sliced into one-second GoPs. The FoV and pFoV are set to 3$\times$3 tiles. The feasible set of the streamed zone is \{3$\times$3, 3$\times$5, 3$\times$7, 4$\times$7, 4$\times$8\}. The set is determined by initially ensuring the shape of the zone is a rectangle, followed by linearly increasing the number of tiles in the zone.
The server first ensures that all tiles within the zone are streamed with the lowest quality representation. When extra transmission resources are available, the server first increases the quality of central tiles in pFoV to the highest quality, then increases the quality of the remaining tiles in pFoV to the highest level, and finally increases the quality of the tiles in the zone but outside the pFoV. If there are still available transmission resources after all tiles within the zone have been increased to the highest quality, the server extends the high-quality streaming to tiles located outside the zone.
The size of the zone depends on the uploaded prediction error. When
the uploaded prediction error is zero, then 3$\times$3 is selected. When the predicted error achieves the maximum, 4$\times$8 is selected.
Assume that the current 5G system can support the streaming of tiles within pFoV at 4K resolution and the remaining tiles at 720p within $T_{\mathrm{ps}}=0.95$~s~\cite{Xing_VR_Shannon,VREXP}.

After playing a VR video, the quality of the video in gaze direction and the remaining region of the actual FoV, quality variation among time, stalling time, and initial delay are measured and reported. The overall QoE for each video $\mathrm{QoE}_s\in[1,5]$ is calculated as a weighted sum of the above metrics \cite{VREXP}. The detailed expression can be found in lemma 1 of \cite{VREXP}.
When $\mathrm{QoE}=5$, it means that the tiles in the FoV are always displayed in the highest quality representation, the stalling time is zero, and the initial delay is less than 0.1 s \cite{VREXP}.
The value of QoE is first calculated for each video and then averaged over the testing set.

The performance gain of B-PEA is evaluated by comparing with the following baselines. Adding Gaussian noise~\cite{privacy-preserving_eye_tracking_2021,LX19,SHIM+19} and Laplace noise~\cite{nair22going} to the actual viewpoint, which are with legends ``Gaussian" and ``Laplace", respectively.\footnote{We also consider adding Gaussian and Laplace noises to the predicted viewpoint, but the results are similar to adding these noises to the actual viewpoint in terms of satisfying privacy requirements and QoE-privacy tradeoff. For conciseness, these results are not shown.}
The mean values of noises for the baselines are set as zero. The standard deviation of Gaussian noise $\sigma$ and the scalar factor of Laplace noise $b$ are set according to the privacy requirement.

\subsection{Privacy Preservation}
To gain useful insight, we consider that all users have identical privacy requirement $q\in[0,1]$.
To obtain the policies of the baselines, we conduct a one-dimensional search to find the minimum $\sigma$ and $b$ that satisfy $\mathrm{Pr}(A)\leq q$ over the training set. For B-PEA, the amount of noise can be directly obtained by substituting the value of $q$, $e$, and $\varepsilon$ into the solution of P2.
We then evaluate the performance of each approach in terms of privacy requirement satisfaction ratio (PSPR) over the testing set.
The PSPR is the percentage of the users whose privacy requirement are satisfied over the whole set, i.e.,
\begin{align}
\mathrm{PSPR}\triangleq \frac{\sum_{j=1}^{N_{\mathrm{test}}} \sum_{k=1}^K \mathbbm{1}(\mathrm{Pr}(A)_{j,k}\leq q)}{N_{\mathrm{test}}K}, \nonumber
\end{align}
where $\mathrm{Pr}(A)_{j,k}$ is the viewpoint leakage probability of the $k$-th user watching the $j$-th video.

\begin{figure}
	\centering
	\begin{minipage}[t]{1\linewidth}
		\includegraphics[width=1\textwidth]{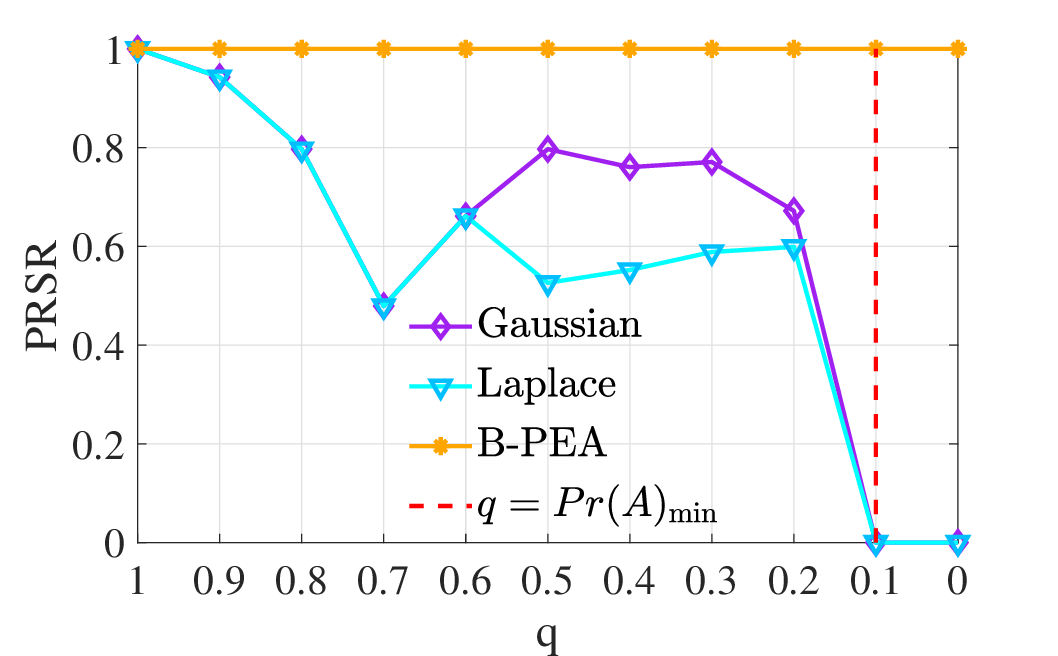}
	\end{minipage}
	\caption{PRSR v.s. $q$}
	\label{Fig:PRSR_q}
	\vspace{-0.4cm}
\end{figure}

In Fig. \ref{Fig:PRSR_q}, we compare the PRSR of the three approaches. we can observe that B-PEA can satisfy the privacy requirement for all users, while the baselines cannot. Besides, the baselines fail to satisfy the privacy requirement for arbitrary users when $q\leq \mathrm{Pr}(A)_{\min}$, which is the minimal leakage probability given by \eqref{lemma:min_mutual_information}. These results validate the analysis in Sec.~III-B.

\subsection{QoE-Privacy Tradeoff}
We next evaluate the prediction performance-privacy tradeoff and the QoE-privacy tradeoff, which are obtained by averaging prediction errors and QoE as a function of the achievable viewpoint leakage probability over the testing set.

The baselines cannot satisfy the privacy requirement. To reduce the achievable viewpoint leakage probability, we increase the standard deviation of Gaussian noise $\sigma$ and the scale factor of Laplace noise $b$ until the probability remains unchanged, where $\sigma\in[0,7]$ and $b\in[0,6]$. For B-PEA, we set $q\in[0,0.7]$, where $0.7$ is the viewpoint leakage probability without noise.

\begin{figure}
	\centering
	\subfloat[Prediction errors v.s. achievable viewpoint leakage probability.]{\label{Fig:PE_Gau_E}
		\begin{minipage}[c]{0.9\linewidth}
			\centering
			\includegraphics[width=1\textwidth]{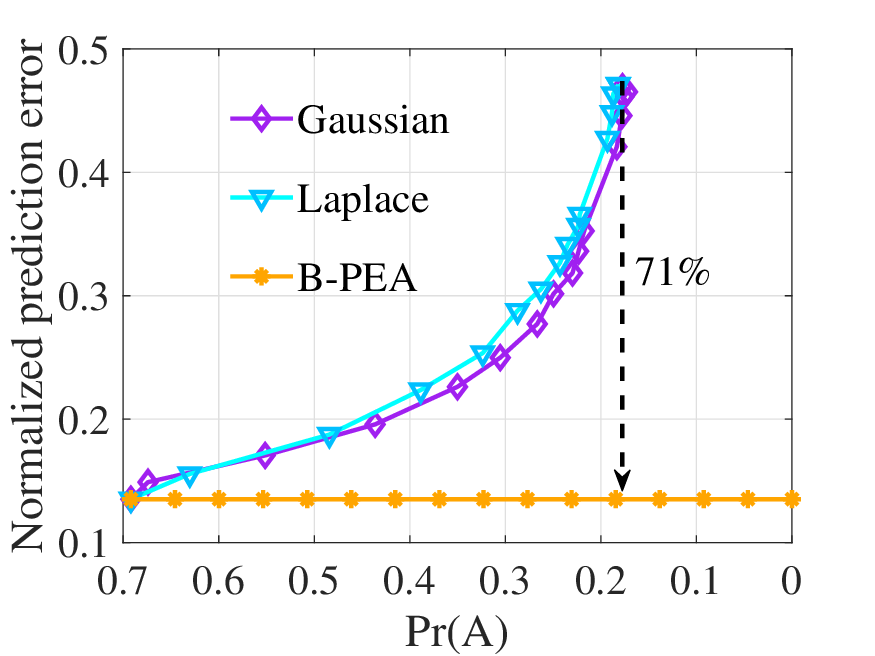}
		\end{minipage}
	}\\
	\subfloat[QoE v.s. achievable viewpoint leakage probability]{\label{Fig:PE_Cst_E}
		\begin{minipage}[c]{0.9\linewidth}
			\centering
			\includegraphics[width=1\textwidth]{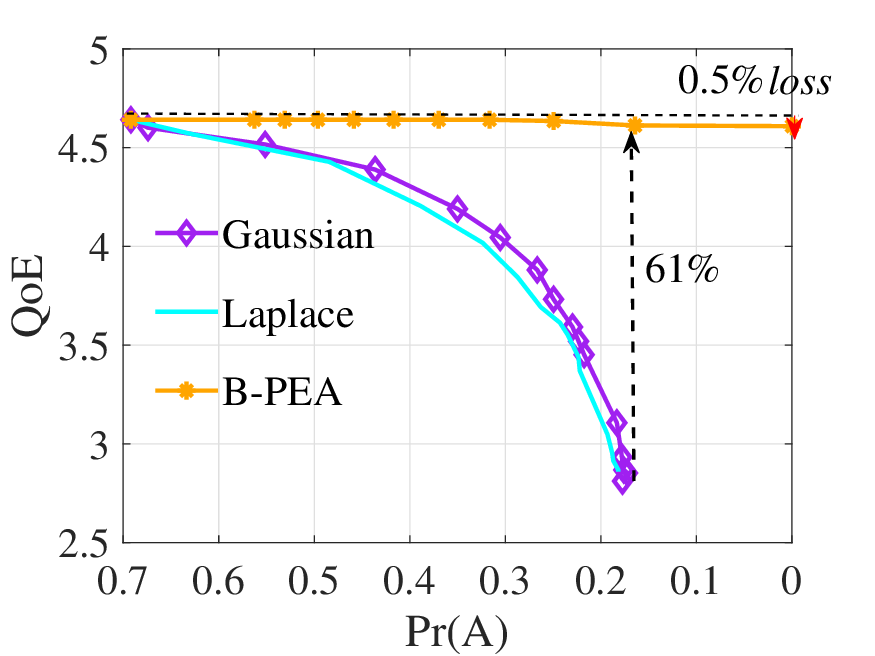}
		\end{minipage}
	}
\captionsetup{font=small, labelsep=period}
	\caption{Comparison on prediction performance-privacy tradeoff and QoE-privacy tradeoff.}\label{Fig:PE_QoE-privacy_tradeoff}	
	\vspace{-0.4cm}
\end{figure}

Figure \ref{Fig:PE_Gau_E} shows the average viewpoint prediction errors, normalized by dividing $\pi$, achieved by B-PEA and the baselines. As the viewpoint leakage probability decreases, we observe that prediction errors for the baselines increase rapidly, while there is no loss of prediction performance for B-PEA. Specifically, when the achievable viewpoint leakage probability is 0.18, the resulting prediction errors of B-PEA are reduced by 71\% compared to the baselines.

Figure \ref{Fig:PE_Cst_E} shows the average QoE versus the achievable viewpoint leakage probability. As expected, B-PEA outperforms the baselines in terms of QoE. Specifically, when the achievable viewpoint leakage probability is 0.18, the QoE for B-PEA is improved by 61\% compared to the baselines. Besides, we observe that the QoE loss of B-PEA is marginal. To achieve zero viewpoint leakage probability, the QoE loss is merely 0.5\%.

\section{Conclusion}\label{section:conclusion}

This paper analyzed and optimized the QoE-privacy tradeoff for proactive VR streaming. For existing privacy-preserving approaches, we derived the viewpoint leakage probability, and obtained the optimal distribution of prediction errors that minimizes the leakage probability. The results revealed that existing approaches cannot fully mitigate viewpoint leakage. Then, we proposed the B-PEA approach, which can achieve zero viewpoint leakage by adding noise to the uploaded prediction errors. We first derived the viewpoint leakage probability for B-PEA, and then obtained the optimal noise-adding strategy that minimizes the amount of noise while satisfying the requirement of the viewpoint leakage probability. Simulation results using a SOTA viewpoint predictor and a practical VR streaming platform validated our performance analysis, and showed that the proposed approach can dramatically reduce the loss of QoE compared to the baselines under the same requirements of viewpoint leakage probability.

\begin{appendices}
\numberwithin{equation}{section}
\section{Proof of Proposition 1}\label{appendix:Prop_1}

\begin{figure}
	\centering
	\subfloat[$\theta\leq e - \varepsilon$]{\label{Fig:V_Pr_0_small_e}
		\begin{minipage}[c]{0.44\linewidth}
			\centering
			\includegraphics[width=1\textwidth]{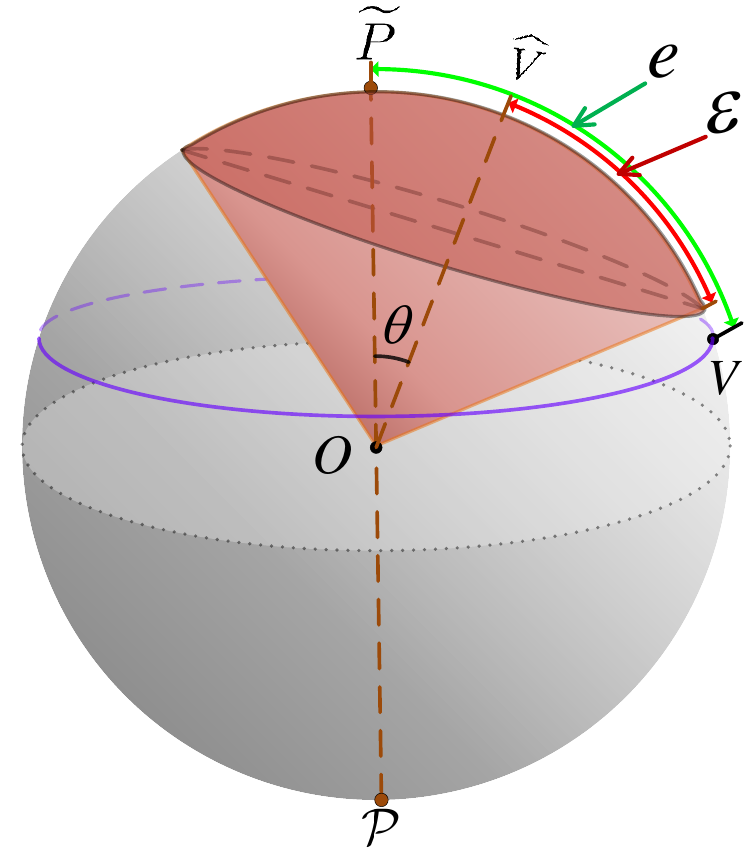}
		\end{minipage}
	}
	\subfloat[$\theta \geq e + \varepsilon$]{\label{Fig:V_Pr_0_large_e}
		\begin{minipage}[c]{0.44\linewidth}
			\centering
			\includegraphics[width=1\textwidth]{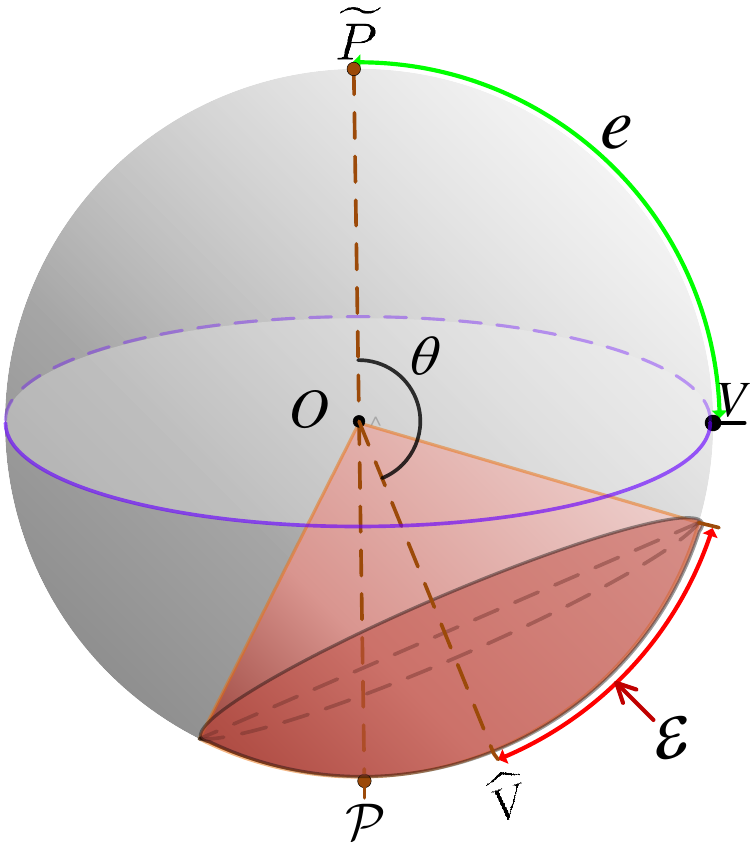}
		\end{minipage}
	}
\captionsetup{font=small, labelsep=period}
	\caption{Examples of $\widehat{V}$ when $\theta\leq e -\varepsilon$ or $\theta\geq e + \varepsilon$}\label{Fig:V_Pr_0}	
	\vspace{-0.4cm}
\end{figure}

To find the optimal $\widehat{V}$ for $e\in[\varepsilon,\pi - e]$, let $\theta$ denote the angle between the line $\overline{\widetilde{P} O}$ and the line $\overline{\widehat{V} O}$, as shown in Fig. \ref{Fig:V_Pr_0}. The optimal $\theta$ can maximize the leakage probability, i.e., $\theta^*=\arg\max_{\theta\in[0,\pi]}\mathrm{Pr}(Q|\widetilde{P}, e, \widehat{V}(\theta))$

As shown in Fig. \ref{Fig:V_Pr_0}, when $\theta< e -\varepsilon$ or $\theta> e + \varepsilon$, the neighborhood cannot include $V$, leading to $\mathrm{Pr}(Q|\widetilde{P}, e,\widehat{V}(\theta))=0$. Thus, the optimal $\theta^*$ should lies in the range $\theta\in[e -\varepsilon, e + \varepsilon]$.

\begin{figure}
	\centering
	\subfloat[$\theta\geq e$]{\label{Fig:Optimal_V_maincase_1}
		\begin{minipage}[c]{0.44\linewidth}
			\centering
			\includegraphics[width=1\textwidth]{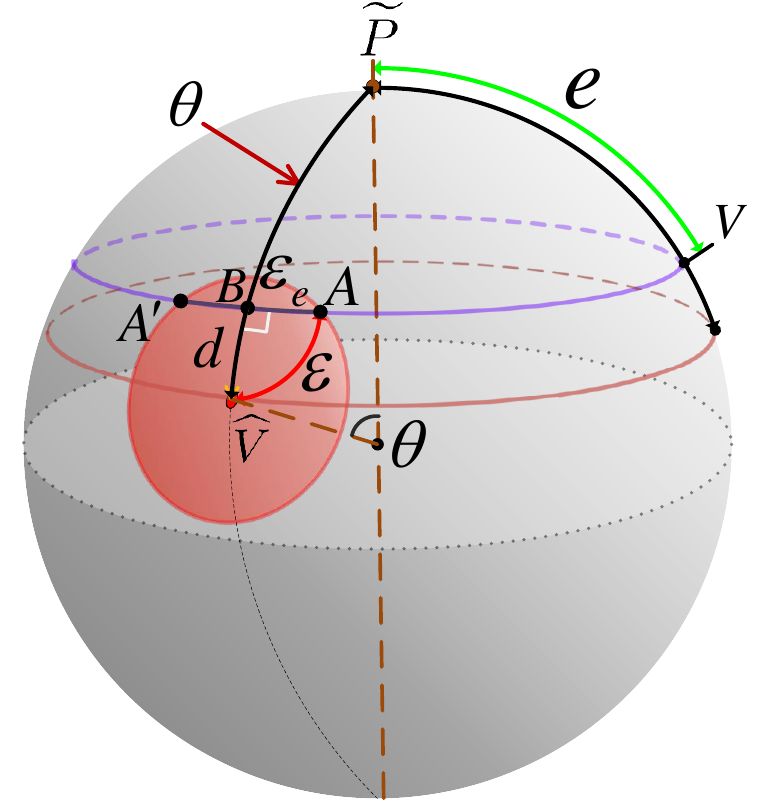}
		\end{minipage}
	}
	\subfloat[$\theta\leq e$]{\label{Fig:Optimal_V_maincase_2}
		\begin{minipage}[c]{0.44\linewidth}
			\centering
			\includegraphics[width=1\textwidth]{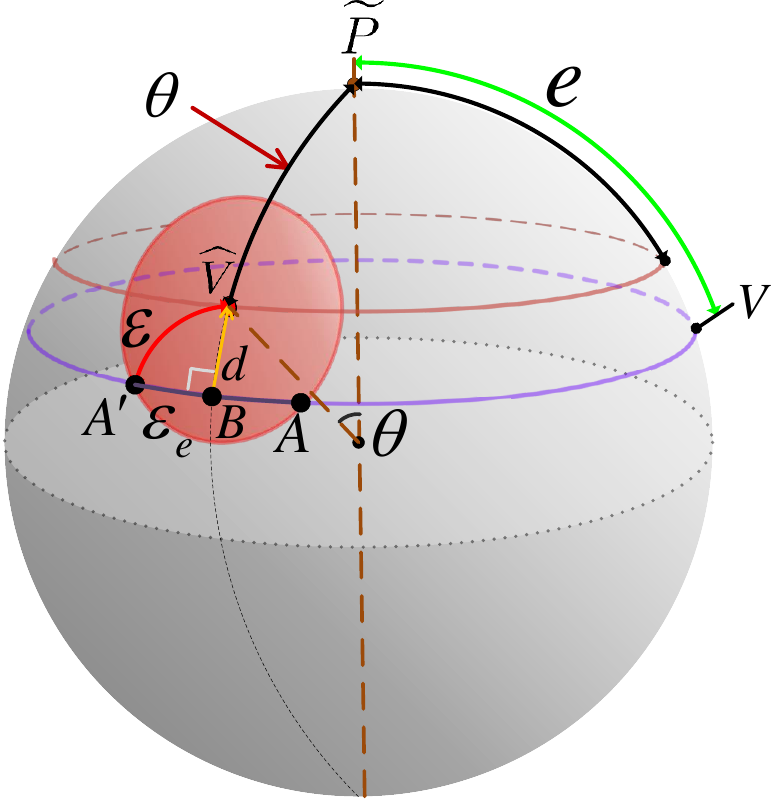}
		\end{minipage}
	}
\captionsetup{font=small, labelsep=period}
	\caption{Examples of $\widehat{V}$ when $\theta\in[e -\varepsilon, e + \varepsilon]$}\label{Fig:Optimal_V_maincase}	
	\vspace{-0.4cm}
\end{figure}

When $\theta\in[e, e + \varepsilon]$, as shown in Fig. \ref{Fig:Optimal_V_maincase_1}, the length of the arc $\bigfrown{\widetilde{P}\widehat{V}}$ equals to $\theta$ when $\theta$ is measured in radians. Then, we obtain the distance $d=\theta-e$ as defined in Fig. \ref{Fig:Optimal_V_maincase}.
Points $\widehat{V}$, $A$, and $B$ form a spherical right triangle. According to the spherical Pythagorean theorem, we have $\cos\varepsilon=\cos\varepsilon_e\cos(\theta-e)$, from which we obtain $\varepsilon_e=\arccos\left[\frac{\cos\varepsilon}{\cos(\theta-e)}\right]$, where the distance $\varepsilon_e$ is defined in Fig. \ref{Fig:Optimal_V_maincase}. We can observe that $\varepsilon_e$ increases as the decrease of $\theta$, since $\theta\in[e, e + \varepsilon]$ and $\varepsilon<0.5\pi$ as we considered.
As $\varepsilon_e$ increases, more parts of the circle is included in the neighborhood, and thus $\mathrm{Pr}(Q|\widetilde{P}, e,\widehat{V}(\theta))$ increases. Therefore, $\mathrm{Pr}(Q|\widetilde{P}, e,\widehat{V}(\theta))$ is maximized when $\theta$ reaches the minimal value, i.e., $\theta^*=e$. Similarly, when $\theta\in[e -\varepsilon, e]$, as shown in Fig. \ref{Fig:Optimal_V_maincase_2}, we can obtain that $\theta^*=e$. Combining the two cases, we obtain that $\theta^*=e$ for $\theta\in[e -\varepsilon, e + \varepsilon]$. The inferred points $\widehat{V}$ corresponding to $\theta^*$ form a circle where the actual viewpoint $V$ is located.


\section{Optimal Solution to Problem \eqref{lemma:functional_prob}}\label{appendix}
The augmented variational problem for the constrained variational problem \eqref{lemma:functional_prob} is
\begin{align}
    \mathcal{L} =& \int_{0}^{\pi} f_e(e)\mathrm{Pr}(A|\widetilde{P}, e)\mathrm{d}e  - \lambda \left(\int_{0}^{\pi}f_e(e)\mathrm{d}e-1\right)\nonumber\\
    \triangleq& \int_{0}^{\pi} \left[S(e,f_e(e))\right]\mathrm{d}e + \lambda,
\end{align}
where $\lambda$ is the Lagrange multiplier, and
$S(e,f_e(e))\triangleq f_e(e)\mathrm{Pr}(A|\widetilde{P}, e) - \lambda f_e(e)$.

According to Euler-Lagrange equation~\cite{fox1987introduction}, the optimal $f_e(e)$ satisfies $\frac{\partial S(e,f_e(e))}{\partial f_e(e)}=\frac{d}{d e}\frac{\partial S(e,f_e(e))}{\partial f'(e)}$, where $f'(e)$ is the first-order derivative of $f_e(e)$. Since $S(e,f_e(e))$ is unrelated to $f'(e)$, $\frac{\partial S(e,f_e(e))}{\partial f'(e)}=0$. Then, we can obtain from the Euler-Lagrange equation that
\begin{align}
    \frac{\partial S(e,f_e(e))}{\partial f_e(e)}
\!=\!\left
\{\begin{array}{lr}
1 - \lambda=0, \ \ \ \ \ \ \ \ \ \ e\leq\varepsilon \ \textrm{or} \ e\geq\pi - \varepsilon,  \\
\min\left(\frac{\varepsilon}{\pi \sin e},1\right)- \lambda=0, \ e\in[\varepsilon, \pi - \varepsilon],
\end{array}
\right.
\label{lemma:variation_Lagrangian_function}
\end{align}

For the case where $e\leq \varepsilon$ or $e\geq\pi-\varepsilon$, we find that the Euler-Lagrange equation holds for arbitrary $f_e(e)$. In this case, we can obtain from \eqref{E:I1} that $\mathrm{Pr}(A)=1$.

For the case where $e\in[\varepsilon,\pi-\varepsilon]$, we can obtain from \eqref{lemma:variation_Lagrangian_function} that
\begin{align}
e = \arcsin\left(\frac{\varepsilon}{\lambda\pi}\right)\triangleq e_0
\end{align}
Further considering \eqref{lemma:functional_cst}, we can find that the optimal $f_e(e)$ is the Dirac delta function, i.e., $f^*(e)=\delta(e - e_0)$.
By substituting $f^*(e)$ into \eqref{E:I1}, we obtain that $\mathrm{Pr}(A)=\frac{\varepsilon}{\pi\sin e_0}$, whose minimal value is $\frac{\varepsilon}{\pi}$, which is achieved when $e_0=0.5\pi$.

By comparing the values of $\mathrm{Pr}(A)$ in the two cases, we obtain that $f^*(e)=\delta(e - e_0)$.

\section{Optimal solution to Problem \eqref{lemma:functional_BPEA_prob}}\label{appendix:optimal_f_e}
The augmented variational problem for the constrained variational problem \eqref{lemma:functional_BPEA_prob} is
\begin{align}
    \mathcal{L} &= \int_{e} f_e(e) \int_{n} |n| f_{n|e,\varepsilon}(n) \mathrm{d} n \mathrm{d} e - \mu\left(\int_{n} f_{n|e,\varepsilon}(n)\mathrm{d} n - 1\right)\nonumber \\
    & \ \ \
    -\lambda \left(\int_{e}f_e(e)\int_{n} f_{n|e,\varepsilon}(n) \mathrm{Pr}(A|\widetilde{P}, e, n) \mathrm{d} n  \mathrm{d} e - q \right) \nonumber\\
    &\triangleq \int_{e} f_e(e) \int_{n} R(n,f_{n|e,\varepsilon}(n)) \mathrm{d} n  \mathrm{d} e   + \mu - \lambda q
\end{align}
where $\lambda$ and $\mu$ are the Lagrange multipliers, and $R(n,f_{n|e,\varepsilon}(n))\triangleq |n| f_{n|e,\varepsilon}(n) - \mu f_{n|e,\varepsilon}(n) - \lambda f_{n|e,\varepsilon}(n)\mathrm{Pr}(A|\widetilde{P}, e, n)$.

According to Euler-Lagrange equation\cite{fox1987introduction}, the optimal $f_{n|e,\varepsilon}(n)$ satisfies $\frac{\partial R(n,f_{n|e,\varepsilon}(n), e, \varepsilon)}{\partial f_{n|e,\varepsilon}(n)}=\frac{d}{d n}\frac{\partial R(n,f_{n|e,\varepsilon}(n), e, \varepsilon)}{\partial f'(n)}$, where $f'(n)$ is the first-order derivative of $f_{n|e,\varepsilon}(n)$. Since $R(n,f_{n|e,\varepsilon}(n), e, \varepsilon)$ is unrelated to $f'(n)$, $\frac{\partial R(n,f_{n|e,\varepsilon}(n), e, \varepsilon)}{\partial f'(n)}=0$. Then, we can obtain from the Euler-Lagrange equation that
\begin{align}
   & \frac{\partial R(n,f_{n|e,\varepsilon}(n))}{\partial f_{n|e,\varepsilon}(n)}\nonumber\\
   &= |n| - \mu - \lambda \int_{e} f_e(e) \mathrm{Pr}(A|\widetilde{P}, e, n) \mathrm{d} e = 0.\label{lemma:variation_Lagrangian_function_BPEA}
\end{align}

The Euler-Lagrange equation \eqref{lemma:variation_Lagrangian_function_BPEA} should hold for any $n$.
Letting $n=\mu$ or $n=-\mu$, we can obtain from \eqref{lemma:variation_Lagrangian_function_BPEA} that
$\lambda \int_{e} f_e(e) \mathrm{Pr}(A|\widetilde{P}, e, \mu) \mathrm{d} e = 0$ and $\lambda \int_{e} f_e(e) \mathrm{Pr}(A|\widetilde{P}, e, -\mu) \mathrm{d} e = 0$. We can find from Table \ref{table:I_e_eps_n} that $\mathrm{Pr}(A| \widetilde{P}, e, \mu)$ and $\mathrm{Pr}(A| \widetilde{P}, e, -\mu)$ cannot be zeros simultaneously. Therefore, we obtain that $\lambda=0$ and $n$ has a deterministic value, which is either $\mu$ or $-\mu$. Further considering \eqref{lemma:functional_BPEA_cst2}, we can find that the optimal $f_{n|e,\varepsilon}(n)$ is the Dirac delta function.

\end{appendices}

\bibliographystyle{IEEEtran}
\bibliography{IEEEabrv_Xing,ref}

\end{document}